\documentclass[11pt,Chicago]{amsart}
\usepackage{amssymb,graphicx,amsmath,amsfonts,multirow,lscape,afterpage,caption,color}
\usepackage{bibentry}
\nobibliography*
\usepackage{lineno}
\usepackage{lscape}
\usepackage{flafter}
\usepackage{afterpage}
\usepackage{enumitem}
\usepackage{pdfpages}
\newcommand{\idays}{days$^{-1}$}
\newcommand{\icells}{cells$^{-1}$}
\newcommand{\mcells}{*10$^6$cells}
\newcommand{\betacells}{$\beta$-cells}
\newcommand{\betacell}{$\beta$-cell}
\newcommand{\IFNg}{IFN-$\gamma$}

\author{James R. Moore}
\address{Department of Mathematics, University of Utah, 155 S 1400 E Rm 233, Salt Lake City, UT 84111, United States}
\curraddr{School of Mathematics, Georgia Institute of Technology, 686 Cherry Street, Atlanta, GA 30332, United States}
\email{jmoore323@math.gatech.edu}

\author{Fred Adler}
\address{Department of Mathematics, University of Utah, 155 S 1400 E Rm 233, Salt Lake City, UT 84111, United States}
\email{adler@math.utah.edu}
\title{Mathematical modeling of Type 1 diabetes progression in the NOD mouse: separating incidence and age of onset.}
\begin{document}
\begin{abstract}
Type 1 diabetes (T1D) is an autoimmune disease of the \betacells{} of the pancreas. The nonobese diabetic (NOD) mouse is a
commonly used animal model, with roughly an 80\% incidence rate of T1D among females. In 100\% of NOD mice, 
macrophages and T-cells invade the islets in a process called insulitis. It can be several weeks between insulitis and T1D, and some
mice do not progress at all. It is thought that this delay is mediated by regulatory T-cells (Tregs) and that a gradual loss of effectiveness in
this population leads to T1D. However, this does not explain why some mice progress and others do not.
We propose a simple mathematical model of the interaction between \betacells{} and the immune
populations, including regulatory T-cells. We find that individual mice may enter one of two stable steady states: a `mild' insulitis state that does not progress
to T1D and a `severe' insulitis state that does. We then run a sensitivity analysis to identify which parameters affect incidence of T1D versus those that
affect age of onset. We also test the model by simulating several experimental manipulations found in the literature that modify insulitis severity and/or
Treg activity. Notably, we are able to match a reproduce a large number of phenomena using a relatively small number of equations. We finish by proposing experiments that
could help validate or refine the model.
\end{abstract}
\maketitle
\section{Introduction}
\subsection{Biological background}
\begin{figure}[htbp]
	\includegraphics[width=\textwidth]{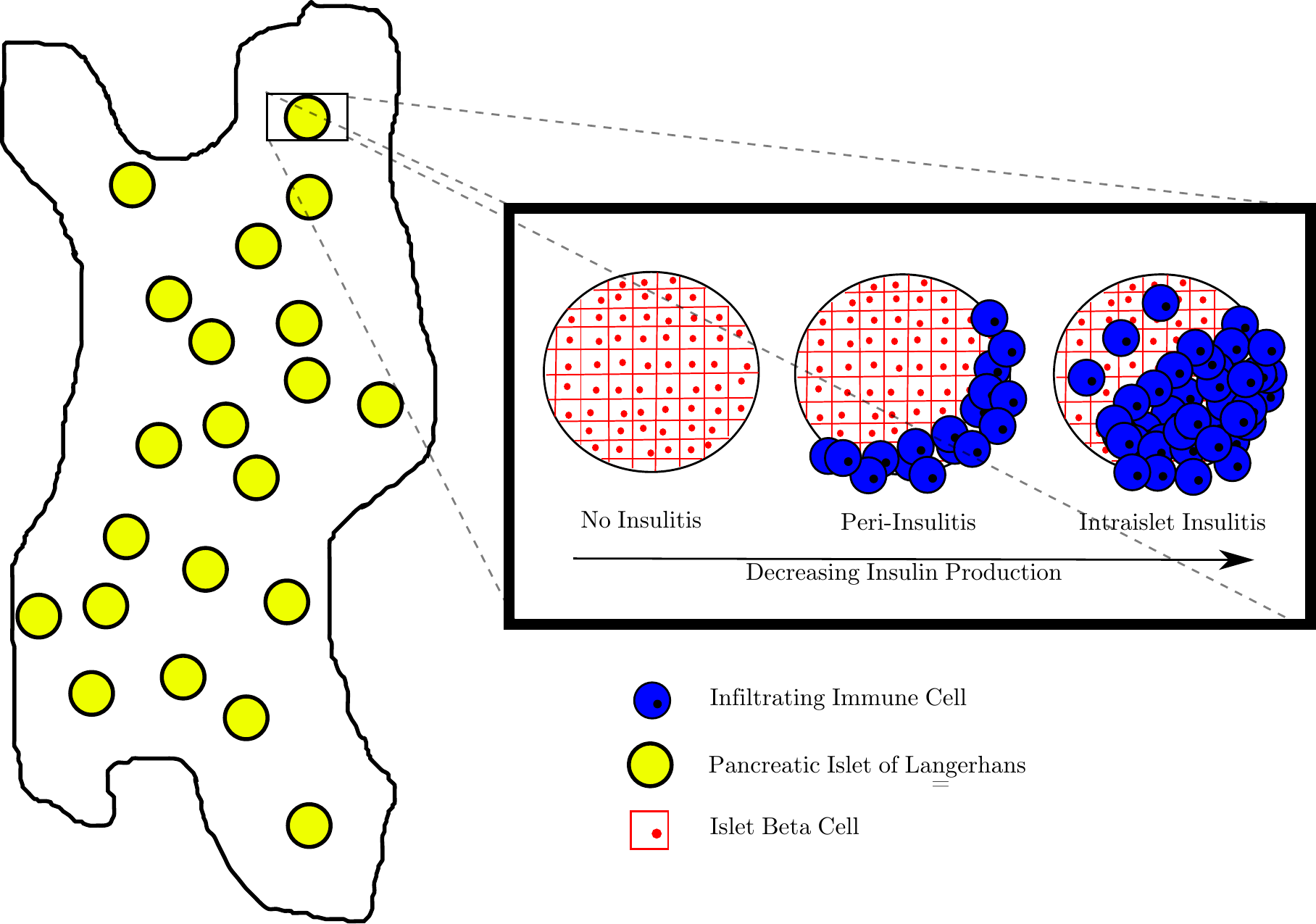}
	\caption{Progression of insulitis in the islets. Over time, immune cells such as T-cells and macrophages infiltrate
the islets, killing \betacells{} and decresing insulin production.}
	\label{fig:insulitis}
\end{figure}

Type 1 diabetes (T1D) is an autoimmune disorder in which T-cells invade the
islets of langerhans within the pancreas and kill insulin producing \betacells{}.
The nonobese diabetic or NOD mouse is an inbred mouse strain that spontaneously develops
T1D. Among females, the age of onset is on average 12-16 weeks, with an incidence of
60-80\%. Prior to T1D onset, at 3-4 weeks of age, immune cells such as CD4 T-cells,
CD8 T-cells, and macrophages invade the islets. This infiltration, called \textbf{insulitis}, gradually becomes more severe,
affecting more islets and penetrating more deeply (see Figure \ref{fig:insulitis}). It is present
in all NOD mice, even those that do not develop T1D. 

NOD mice exhibit multiple
immune problems. For an extensive review, see \cite{anderson2005nod}.
Here we focus on a few key differences from the wild type, which may mimic
the factors that lead to genetic susceptibility in humans.
First, there is a \textbf{failure in central tolerance} as NOD mice do not
effectively present peptides from the proinsulin gene in the thymus.
This leads to the generation of a population of proinsulin reactive
lymphocytes, although these T-cells typically have a low affinity for their
target T-cell antigen.
These T-cells are the first detected in the pancreas during immunopathogenesis
and eventually give way to other more reactive T-cell clones. However, the anti-proinsulin
response appears to be required \cite{french1997transgenic, zhang1991suppression}.
In humans, T1D is associated with lower expression of insulin in the thymus \cite{jayasimhan2014advances}.
Second, NOD mice have a \textbf{defect in the clearance of apoptotic cells}.
Specifically, this is a defect in \textbf{macrophages}, an antigen nonspecific immune
cell with a large number of jobs, including the clearance of dead cells and the activation
of T-cells. The excess apoptotic cells can
become necrotic and trigger an inflammatory response by macrophages. This defect is particularly important
at the time of weaning when the pancreas undergoes structural changes and heightened apoptosis \cite{trudeau2000neonatal}. 
This \textbf{apoptotic wave} and subsequent inflammation initiates an immune response, and T-cells
start to infiltrate the pancreas (see Table \ref{events}).
Finally, NOD mice have a \textbf{defect in the growth and survival factor IL-2}. Despite its role as a T-cell
growth factor, deficiency in IL-2 typically leads to uncontrolled growth of effector T-cells as IL-2 is required
 for the proliferation and survival of Tregs. Tregs can control the development
 of T1D in the NOD mouse for several months, and T1D is greatly accelerated in Treg-deficient
 NOD mice \cite{chen2005where, feuerer2009punctual}. Human T1D patients have impaired
IL-2 signaling via a defect in the high affinity IL-2 receptor \cite{jayasimhan2014advances}.

\begin{table}[phtb]
\begin{center}
\caption{T1D progression in the NOD mouse}
\begin{tabular}{lll}
\hline
Time & Event & Source\\
\hline
9-15 days & Apoptotic Wave &\cite{trudeau2000neonatal}\\
18 days & T1D in Treg deficient NOD mice &\cite{chen2005where}\\
3 weeks & Initiation of Insulitis &\cite{fu2012early}\\
4-5 weeks & Insulin-specific CD8s dominate & \cite{trudeau2003prediction}\\
4-5 weeks & Rapid T1D after ablation of Tregs & \cite{feuerer2009punctual}\\
6 weeks & Differential Prognosis with MRI & \cite{fu2012early}\\
6-8 weeks & Decline in Treg effectiveness & \cite{tritt2008functional}\\
8 weeks & IGRP-specific CD8s appear & \cite{trudeau2003prediction}\\
8-12 weeks & \betacell{} mass starts to decline & \cite{alanentalo2010quantification}\\
12-16 weeks & Rapid loss of \betacell{} mass & \cite{alanentalo2010quantification}\\
16 weeks & Median T1D onset & \cite{amrani2000progression}\\
\hline
\end{tabular}
\label{events}
\end{center}
\end{table}

It is unclear what causes T-cells to escape the regulation of Tregs and destroy \betacells{}. The simplest
hypothesis is that the destruction of \betacells{} within the islets is ongoing, but that T1D is not diagnosed
until the \betacell{} mass reaches a critical level. However, quantification of the \betacell{} mass shows that
it does not start to decline until 8-12 weeks of age \cite{alanentalo2010quantification}, 6-10 weeks after the apoptotic
wave \cite{trudeau2000neonatal}. Ablation of Tregs in 4-6-week old mice leads to rapid T1D onset \cite{feuerer2009punctual},
indicating that Tregs are required to prevent disease progression. This suggests that Tregs possibly lose
effectiveness over time. This hypothesis is supported by the work of Tritt et al.\ \cite{tritt2008functional},
who find that older mice have similar numbers of Tregs, but they are less able to control T1D than those of young adult mice, and
Pop et al.\ \cite{pop2005single}, who find that Tregs from older mice have a loss of function \textit{in vitro}.
Another possibility is that \betacells{} gradually lose function over time. 
It may be that the \betacells{} degranulate, losing their ability to produce insulin, 
\cite{akirav2008beta} or simply apoptose \cite{graham2012mathematical, topp2000model}
in response to increased demands. Finally, the delay may be due to the time it takes for the development
of a population of T-cells capable of killing \betacells{}. In the early stages of insulitis,
the most important population of T-cells is \textbf{CD4s}, or helper T-cells, which can activate other components of the immune
sytem, but not directly kill target cells. Another population of T-cells, \textbf{CD8s} or killer T-cells, is very
efficient at killing target cells, but require additional activation. In the early stages of T1D in NOD mice, the CD8 population is generally insulin-specific
with a low affinity. After several weeks, high affinity CD8s, specific to 
islet-specific glucose-6-phosphate catalytic subunit related protein (IGRP), take their place. The
destruction of \betacells{} corresponds to the expansion of this population \cite{amrani2000progression}.

Almost all NOD mice develop insulitis, and yet many do not develop T1D.
Trudeau et al.\ \cite{trudeau2003prediction}
find signficant differences in the makeup of the CD8 T-cell population between those that get T1D and those that do not. 
In particular, CD8 T-cells specific for
the islet antigen IGRP are at much higher
levels in the `prediabetic' mice, although they do not appear until week 8 in either group \cite{trudeau2003prediction}. Fu et al.\ \cite{fu2012early} perform MRIs of mice at different ages and find that the degree of inflammation
is significantly greater in the mice destined for T1D. At 6 weeks, inflammation of the pancreas is significantly correlated
with the eventual development of T1D.  They also find that the mice with lower inflammation expressed
higher levels of CRIg, a marker of a class of regulatory macrophages. Taken together, these studies suggest that
the eventual fate of an individual mouse is predetermined at the initiation of insulitis. The nature of the insulitis
of each mouse should therefore fall into at least two classes, which can be distinguished by the presence of CRIg-expressing
macrophages. Islets with more severe inflammation have a greater level of \betacell{}
turnover and therefore a greater presentation of IGRP. This ultimately leads to a greater IGRP CD8 response and T1D.

\subsection{Prior modeling}
As discussed above, the initiation of T1D requires a proinflammatory
stimulus, which, in the NOD mouse, likely occurs during weaning.
Nerup and colleagues \cite{nerup1994pathogenesis} propose
a nonmathematical description of T1D initiation, the `Copenhagen Model', which is
not specific to NOD mice.
The stimulating event, such as a virus, causes minor \betacell{} destruction and, more importantly,
releases \betacell{} antigens triggering an immune cascade. De Blasio et al.\ 
\cite{deblasio1999onset} proposed a simple model of 4 ODEs (resting and activated
	macrophages, \betacell{} antigen and T-cells) to reproduce this phenomenon. The model is intentionally
generic and does not describe any particular environmental insult. They find that the activity
of Macrophages is key and that the nature of the T-cell response does not drive inflammation.
Maree et al.\ \cite{maree2006modelling} expanded upon this model and incorporated the ideas of Trudeau \cite{trudeau2000neonatal}: that NOD mice
have a reduced ability to clear apoptotic \betacells{} that then become necrotic. They find
that for NOD mice, the system is bistable; the wave of apoptosis transfers the system
from a resting `healthy' state to a `disease' state.

Regulatory T-cells play an important role of slowing the progression of T1D and other
autoimmune diseases. Alexander et al.\ \cite{alexander2011self} study a generic model of the Treg-controlled
autoimmune disease. They demonstrate
that Tregs may not eliminate  an autoimmune response, but can reduce it
to a subclinical level. Similarly, Magombdze et al.\ incorporate Treg control into the framework
of the Copenhagen/Maree models and find that Tregs cannot eliminate the autoimmune response, but
can reduce its intensity so that the \betacell{} population is barely affected.

Leah Keshet and colleagues have investigated the dynamics of the CD8 population in a series of papers.
The time series of the IGRP CD8 population in \cite{trudeau2003prediction} appears cyclical.
Mahaffy, Keshet et al.\ \cite{mahaffy2007modeling} model this phenomenon using multiple T-cell compartments:
`activated' T-cells can become either effectors and memory cells (which can later become activated upon restimulation).
Their model reproduces the observed cycle and the \betacells{} die off step-wise during each cycle until none remain.
Khadra, Keshet et al.\ \cite{khadra2009role} investigate the competition between low-affinity and high-affinity
CD8s. High-affinity CD8s kill \betacells{}, releasing antigens and perpetuate the immune response, whereas low-affinity
CD8s simply crowd the environment.  This leads, fairly robustly, to a bistable system that has a `healthy' state with
few high-affinity CD8s and a `diseased' state with many.

\subsection{Outline of our approach}
We seek a simple model that
\begin{enumerate}
\item has two possible outcomes, a `T1D' state and an `insulitis but no T1D' state;
\item develops T1D predominantly within a narrow time window; and
\item is initially under the control of a Treg population, which it subsequently escapes.
\end{enumerate}
We propose a model of T1D development that proceeds in two stages. An `initiation' phase corresponding to
the development-driven apoptosis and a `progression' phase describing the increase in the number and reactivity of the 
islet-specific CD8 population. The initiation phase has two distinct outcomes corresponding to
distinct stable equilibria: \textbf{mild-insulitis}, which does not lead to progression, and \textbf{severe-insulitis}, which does. These two states
are characterized by differences in the makeup of the macrophage population, the ratio of infiltrating Tregs to
effector T-cells, and the cytokine milieu of the islet. We show how various treatments can shift the system
from the severe insulitis state to the mild. The `progression' phase is a feedforward model. Activated macrophages
stimulate the growth of the CD8 population, which in turn kill \betacells{} causing a rise in blood glucose. Only in the
severe insulitis state do the activated macrophages promote a sufficient growth in CD8 cells to promote T1D development.

We then validate this model by simulating various treatments of NOD mice found in the literature and comparing
the results. To replicate incidence data, we must have heterogeneity in the mouse population.  We generate this heterogeneity by
changing the initial number of activated macrophages and the death
rate due to CD8s, which we assume are quite variable. The justification for this is that they are the result of complicated
processes (see \cite{maree2006modelling} and \cite{khadra2009role}, respectively) that can have multiple outcomes in
otherwise identical organisms. We add a small amount of noise to all other parameters.
We then perform a sensitivity analysis to investigate which parameters are key for the development of
T1D. Specifically, we are interested in those that lead to an acceleration or delay of T1D versus those that
change the incidence. Finally, we propose some further extensions and some possible experiments to further validate the model.

\section{The initiation model}
\subsection{Activation of macrophages}
Macrophages are among the first cells to infiltrate the islet in NOD mice \cite{jayasimhan2014advances}, likely
in response to an apoptotic wave of \betacell{} death 
during weaning (the `apoptotic wave', see \cite{trudeau2000neonatal}). As explored in \cite{maree2006modelling},
NOD macrophages are inefficient at clearing these apoptotic cells, which leads to further inflammation and \betacell{}
death. Here, we shall assume that the apoptotic wave has just passed, resulting in an initial excess of macrophages.

Tissue-resident macrophages are a type of innate immune cell, meaning non-antigen-specific, with
an incredibly wide array of potential behaviors. They phagocytose other cells, act as antigen presenting cells (APCs) and control the behavior of neighboring cells, both immune and nonimmune, via the release of cytokines. They are critical
in the initiation and continuation of immune responses and yet they can also act as `custodians', clearing away debris from dead cells. These regulatory activities correlate with the expresion of the receptor CRIg
on a macrophages surface. CRIg expression is promoted by the regulatory molecule IL-10 and
inhibited by the inflammatory molecule \IFNg{} as well as other inflammatory molecules such as arachidonic acid \cite{gorgani2011regulation}. Importantly, the expression of CRIg by pancreatic macrophages is negatively correlated with
the progression to T1D \cite{fu2012early}. Both IL-10 and \IFNg{} are \textbf{cytokines}: diffusing, extracellular molecules used for communication between cells, typically
of the immune system.

We assume there are two classes of Macrophage: inflamatory macrophages ($M^*$),
corresponding to low CRIg expression,
which can stimulate the activation and proliferation of T-cells, and regulatory macrophages ($M$), corresponding to high CRIg expression which act primarily as phagocytes. Macrophages can switch back and forth in response to external signals from
cytokines. All macrophages enter the pancreas at a rate $J$ in the regulatory class. 
They activate to become inflammatory at a basal
rate $a_0$, but their activation rate can be greatly enhanced by \IFNg{}. 
Likewise, inflammatory macrophages deactivate at a basal rate $b_0$, and their deactivation rate is enhanced by IL-10.
\begin{align*}
\frac{dM^*}{dt}&=a M-b M^* -\delta M^*\\
\frac{dM}{dt}&=J-a M+b M^*-\delta M\\
a&=a_{\gamma} \frac{I_{\gamma}^2}{k_{\gamma}^2+I_{\gamma}^2}+a_0\\
b&=b_{10}\frac{I_{10}^2}{k_{10}^2+I_{10}^2}+b_0
\end{align*}

\IFNg{} is produced by TH1 effector CD4 T-cells ($T$) in response to IL-12, a product of inflammatory
macrophages. We let $T^*$ denote the population of effector T-cells expressing \IFNg. IL-10 is produced by
CD4 Tregs ($R$). Pancreatic Tregs in NOD mice overexpress IL-10, so we assume that the entire population expresses IL-10
without the need for further stimulation.
\begin{align}
\frac{dT^*}{dt}&=c \frac{I_{12}^2}{k_{12}^2+I_{12}^2}(T-T^*)-eT^*\\
\frac{dI_{\gamma}}{dt}&=\alpha_{\gamma}T^*-\omega_{\gamma}I_{\gamma}\\
\label{gammeq}
\frac{dI_{12}}{dt}&=\alpha_{12}M^*-\omega_{12}I_{12}\\
\frac{dI_{10}}{dt}&=\alpha_{10}R-\omega_{10}I_{10},
\label{teneq}
\end{align}
where $c$ is the rate at which T-cells are activated by IL-12 and $e$ is the rate
at which they revert to resting. Cytokine $i$ is produced at a rate $\alpha_i$ and
decays at rate $\omega_i$.
We assume that all cytokine concentrations equilibriate rapidly. Therefore, we set \eqref{gammeq}-\eqref{teneq} equal to zero, and solve for
the equilibrium concentrations. This yields
the three-dimensional ODE:
\begin{align}
\begin{split}
\frac{dM^*}{dt}&=-\delta_M M^*+\bigl(a_{\gamma} F(\nu_{\gamma} T^{*})+a_0\bigr)M 
-\left(b_{10} F(\nu_{10} R)+b_0\right) M^*\\
\frac{dM}{dt}&=J-\delta_M M-\bigl(a_{\gamma} F(\nu_{\gamma} T^{*})+a_0\bigr)M
+\left(b_{10} F(\nu_{10} R)+b_0\right) M^*\\
\frac{dT^*}{dt}&=c F(\nu_{12} M^*)(T-T^*)-eT^*
\end{split}
\label{MTeqn}
\end{align}
where
\begin{align*}
F(x)&=\frac{x^2}{x^2+1}\\
\nu_i&=\alpha_i /\omega_ik_i.
\end{align*}

\subsection{Analysis of equilibria}
The equilibria of \eqref{MTeqn} are given by
\begin{align*}
T^*&=C(M^*_{eq})T=\frac{c F(\nu_{12}M^*_{eq})}{e+cF(\nu_{12}M^*_{eq})}T\\
M&=\frac{J}{\delta_M}-M^*_{eq}\\
M^*&=M^*_{eq}
\end{align*}
where $M^*_{eq}$ satisfies
\begin{align}
	\begin{split}
M^*_{eq}&=\frac{J}{\delta_D}\frac{A(M^*_{eq},T)}{\delta_D+B(R)+A(M^*_{eq},T)}\\
A(M^*_{eq},T)&=a_{\gamma} F(\nu_{\gamma} T C(M^*_{eq})+a_0\\
B(R)&=b_{10} F(\nu_{10} R)+b_0.
\end{split}
\label{fpeqn}
\end{align}
\eqref{fpeqn} represents a fifth degree polynomial, whose coefficients alternate signs. Therefore
we expect it to have 1, 3, or 5 positive real roots. For any parameter choice, the number
and value of these equilibria will depend on the state variables $T$ and $R$, whose dynamics
are discussed in the next section.

\begin{figure}[htbp]
	\includegraphics[width=\textwidth]{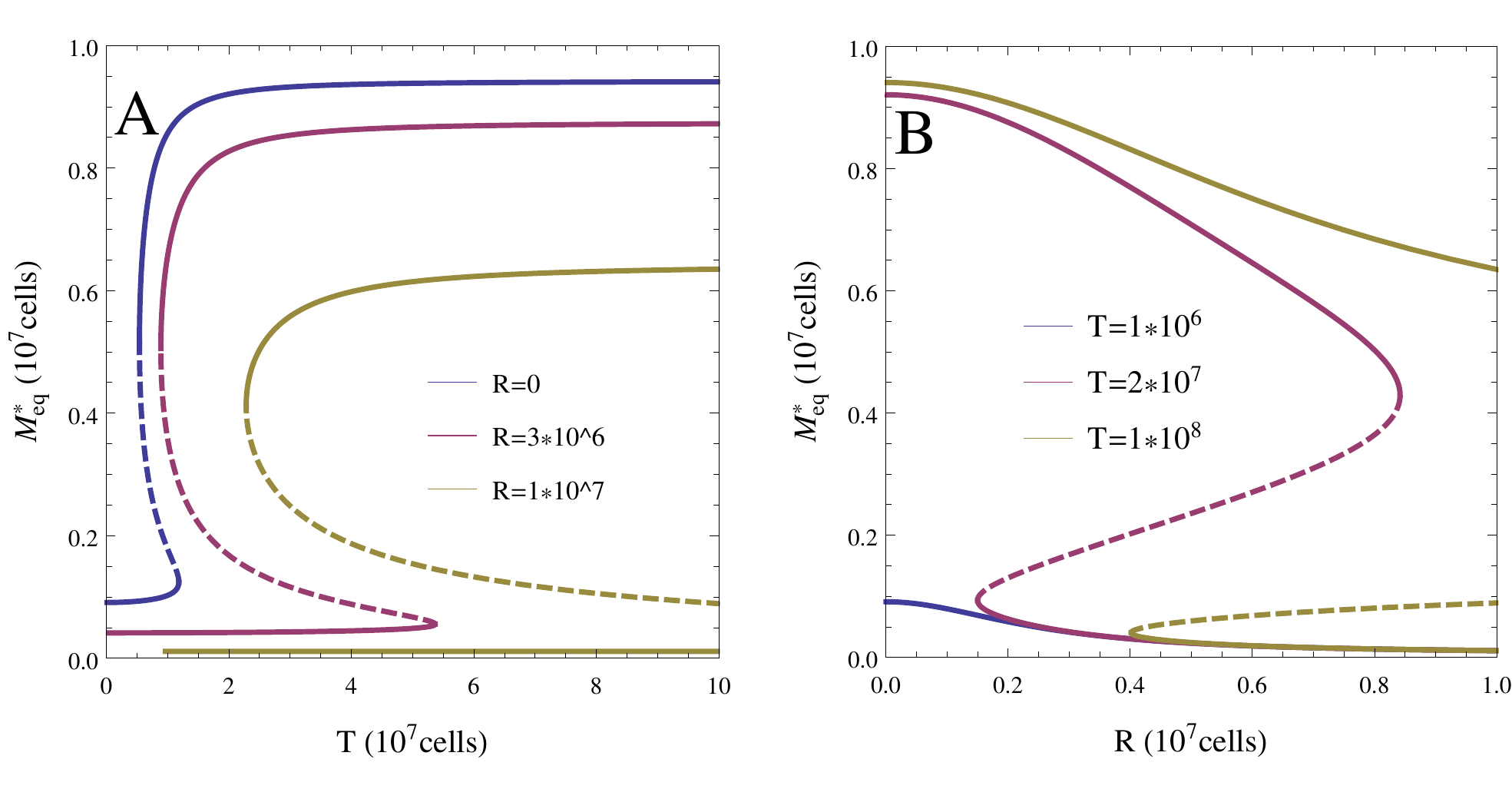}
	\caption{Equilibria of \eqref{MTeqn} with the parameters in Table 3.2.}
	\label{fig:TRL}
\end{figure}

With our parameter values (see Table \ref{dparameters}) there are either
1 or 3 solutions, depending on the values of $T$ and $R$. For fixed $R$ and small values
of $T$, as in Figure \ref{fig:TRL}A, there is only one solution with few inflammatory macrophages.
This agrees with the observation that CD4s are required for the initiation of T1D \cite{anderson2005nod}.
There is also only one solution for large values of $T$, corresponding to large numbers of inflammatory macrophages.
For intermediate values of $T$, there is both the inflamed and noninflamed solutions, separated by an unstable
threshold solution.

\begin{table}[phtb]
\begin{center}
\caption{Parameter values used in the model}
\begin{tabular}{lll}
\hline
Parameter & Description & value\\
\hline
$J$&Influx of macrophages into pancreas&1 \mcells\\
$\delta$ & Turnover of macrophages in pancreas & 1 \idays\\
$a_0$&Basal macrophage activation rate of macrophages&.05 \idays\\
$a_{\gamma}$&Macrophage activation rate induced by IFN-$\gamma$&8 \idays\\
$b_0$&Basal macrophage deactivation rate&.4 \idays\\
$b_{10}$&Macrophage deactivation rate induced by IL-10&8 \idays\\
$c$&CD4 activation rate by IL-12&8 \idays\\
$e$&CD4 deactivation rate&5.5\idays \\
$\nu_{\gamma}$&Scaled IFN-$\gamma$ affinity of macrophages& 2$\delta/J$\\
$\nu_{10}$&Scaled IL-10 affinity of macrophages&$\delta/J$\\
$\nu_{12}$&Scaled IL-12 affinity of CD4 T-cells&$\delta/J$\\

\hline
$\alpha_T$&Influx of effectors into islets&.7 \mcells\\
$\alpha_R$&Influx of Tregs into islets&.3 \mcells\\
$\gamma_T$&Proliferation rate of effectors&2 \idays\\
$\gamma_R$&Proliferation rate of Tregs&2 \idays\\
$\sigma_T$&Max fraction of effectors in mitosis&.4\\
$\sigma_R$&Max fraction of Tregs in mitosis&.8\\
$\delta_{T1}$&Basal death rate of Tregs in pancreas&.2 \idays\\
$\delta_{T2}$&Death rate of Tregs due to IL-2 deficiency&.8 \idays\\
$k_R$&Scaled affinity of Tregs for IL-2&5 $\delta/J$\\
$k_T$&Scaled affinity of effectors for IL-2&20 $\delta/J$\\
\hline
$\alpha_C$&Production of CD8s&.03\\
$\gamma_C$&Division rate of CD8s&.27\\
$\delta_{C1}$&Death rate of CD8s&.01\\
$\delta_{C2}$&Autoregulation rate of CD8s&2.6*$10^{-6}$ \icells\idays\\
$k_C$&Saturation constant for CD8 proliferation&.75 $J$\\
\hline
$\eta_{4}$&Rate of \betacell{} killing by CD4&.3\\
$s_{T4}$&Saturation constant of CD4 killing&1\\
$s_{R4}$&Control of Tregs over CD4 killing&30\\
$\eta_{8}$&Per capita rate of beta cell killing by CD8&3.2 \mcells\\
\hline
$\gamma_B$&Growth rate of beta cells&.06\\
$G_{hb}$&Glucose level for half maximal beta cell growth&161\\
$\delta_B$&Death rate of beta cells&1/60\\
$R_0$&Average glucose production &864 mg\idays\\
$E_{G0}$&Basal glucose decay rate &1.44 \idays\\
$S_I$&Rate of insulin-mediated glucose uptake&.72 \idays/per $\mu U$\\
$\sigma_I$&Max insulin production rate&43.2 $\mu U$\idays per mg \\
$\delta_I$&Insulin decay rate&432 \idays\\
\hline
\end{tabular}
\label{dparameters}

\end{center}
\end{table}

\subsection{Tregs and effectors compete in the islet}
During the progession of T1D, T-cells infiltrate the pancreatic islets.
Cytotoxic CD8 T-cells directly kill $\beta$-cells, but the CD4 population also
plays a role by maintaining an inflammatory environment that perpetuates the immune response.
The CD4 population, also called helper T-cells, is often divided into two broad categories: effector T-cells and regulatory T-cells, which respectively promote or obstruct immune responses. T1D does not develop in NOD mice in the absence
of CD4 effectors. With effectors present but not Tregs, the disease develops much faster, indicating that they also play
an important role \cite{chen2005where}.

Effectors and Tregs enter the islet at a rate of $\alpha_T$ and $\alpha_R$, respectively.
Although there are other populations of antigen presenting cells in the pancreas, we assume that
T-cells are dependent upon interaction with macrophages in order to proliferate. This is equivalent to
assuming that other APC population numbers are correlated with those of inflammatory macrophages.
The maximum per-capita
growth rate should therefore occur when there are very few T-cells relative to macrophages as the competition for
binding space will be minimal. The decrease in proliferation rate as T-cell numbers increase has been observed in vivo \cite{tang2008central}. For simplicity, we model this process as a Michaelis-Menten rate with the inflammatory macrophage acting analagously to an enzyme. We let $\sigma$ denote the maximum proportion of T-cells undergoing mitosis and $\gamma$ denote the exponential growth rate of those dividing cells. The maximum possible growth rate of the T-cell population is therefore $\gamma\sigma$.

The IL-2-BCL-2 pathway controls apoptosis of T-cells in the islets \cite{tang2008central}. Secreted IL-2 promotes the expression
of the antiapoptotic factor BCL-2 in T-cells. Tregs are dependent upon effectors for IL-2, whereas effectors are self sufficient. NOD mice are deficient in IL-2,
leading to increased turnover in Tregs but not effectors \cite{tang2008central}. Therefore, we assume that T-cells die at a constant rate $\delta_{T1}$ but that Tregs will
die at an enhanced rate $\delta_{T1}+\delta_{T2}$ in the absence of IL-2. Effectors secrete IL-2 at a rate $\alpha_2$ and it is taken up
by Tregs at a rate $\omega_2$. In addition, we include an input $u(t)$ representing IL-2 treatment.
\begin{align*}
	\frac{dT}{dt}&=\alpha_T+\gamma_T \frac{\sigma_T M^* T}{M^*+\sigma_T T+\sigma_R R}-\delta_{T1} T\\
	\frac{dR}{dt}&=\alpha_R+\gamma_R \frac{\sigma_R M^* R}{M^*+\sigma_T T+\sigma_R R}-\left(\delta_{T1}+\delta_{T2} \frac{k}{k+I_2}\right) R\\
\frac{dI_2}{dt}&=\alpha_2 T-\delta_2 I_2-\omega_2 I_2 R+u(t)
\end{align*}
Putting the cytokine concentration in steady state,
\begin{align}
\begin{split}
\frac{dT}{dt}&=\alpha_T+\gamma_T \frac{\sigma_T M^* T}{M^*+\sigma_T T+\sigma_R R}-\delta_{T1} T\\
\frac{dR}{dt}&=\alpha_R+\gamma_R \frac{\sigma_R M^* R}{M^*+\sigma_T T+\sigma_R R}-\left(\delta_{T1}+\delta_{T2} \frac{k_R R+1}{Q(t)+k_T T+k_R R +1}\right) R
\end{split}
\label{TReqn}
\end{align}
where $Q(t)=u(t)/k \delta_2$ is the scaled IL-2 treatment, $k_T=\alpha_2/k\delta_2$ the scaled IL-2 production rate and $k_R=\omega_2/\delta_2$ the scaled uptake by Tregs.
Equation \eqref{TReqn} and \eqref{MTeqn} together make up the initiation model.

\subsection{Analysis of the initiation model}
\label{analysissection}
To simplify the analysis of this model, we consider the case $\alpha_T=\alpha_R=0$ with $Q(t)=Q$.
The system \eqref{TReqn} then becomes
\begin{align*}
\begin{split}
0&=\gamma_T \frac{\sigma_T M^*_{eq} T}{M^*_{eq}+\sigma_T T+\sigma_R R}-\delta_{T1} T\\
0&=\gamma_R \frac{\sigma_R M^*_{eq} R}{M^*_{eq}+\sigma_T T+\sigma_R R}-\left(\delta_{T1}+\delta_{T2} \frac{k_R R+1}{Q+k_T T+k_R R +1}\right) R
\end{split}
\end{align*}
which has four solution branches: the trivial solution $T_0=R_0=0$, a `$T$-only' solution with
\begin{align}
T_1&=\frac{\gamma_T \sigma_T-\delta_{T1}}{\sigma_T \delta_{T1}}M_eq\\
R_1&=0
\end{align}
an `$R$-only' solution satisfying
\begin{align}
T_2&=0\\
0&=\gamma_R \frac{\sigma_R M^*_{eq} }{M^*_{eq}+\sigma_R R_2}-\left(\delta_{T1}+\delta_{T2} \frac{k_R R_2+1}{Q+k_R R_2 +1}\right) 
\end{align}
and a `coexistence' solution satisfying
\begin{align}
0&=\gamma_T \frac{\sigma_T M^*_{eq}}{M^*_{eq}+\sigma_T T_3+\sigma_R R_3}-\delta_{T1} \\
0&=\gamma_R \frac{\sigma_R M^*_{eq}}{M^*_{eq}+\sigma_T T_3+\sigma_R R_3}-\left(\delta_{T1}+\delta_{T2} \frac{k_R R_3+1}{Q+k_T T_3+k_R R_3 +1}\right)
\end{align}
Each of these solutions exist for $\alpha_T$,$\alpha_R>0$; however, only the coexistent solution is guaranteed to remain positive.
The other three solutions are positive if and only if they are stable.
Recalling that $M^*_{eq}$ represents the solution of a 5th degree polynomial whose coefficients depend
on $T$ and $R$, we cannot directly solve this system. However, if we view $M^*_{eq}$ as a parameter, we can implicitly
solve for $R$, $T$ and a parameter of our choice to create bifurcation diagrams.

\begin{landscape}
\begin{figure}[phtb]
	\begin{center}
		\includegraphics[width=22.5cm]{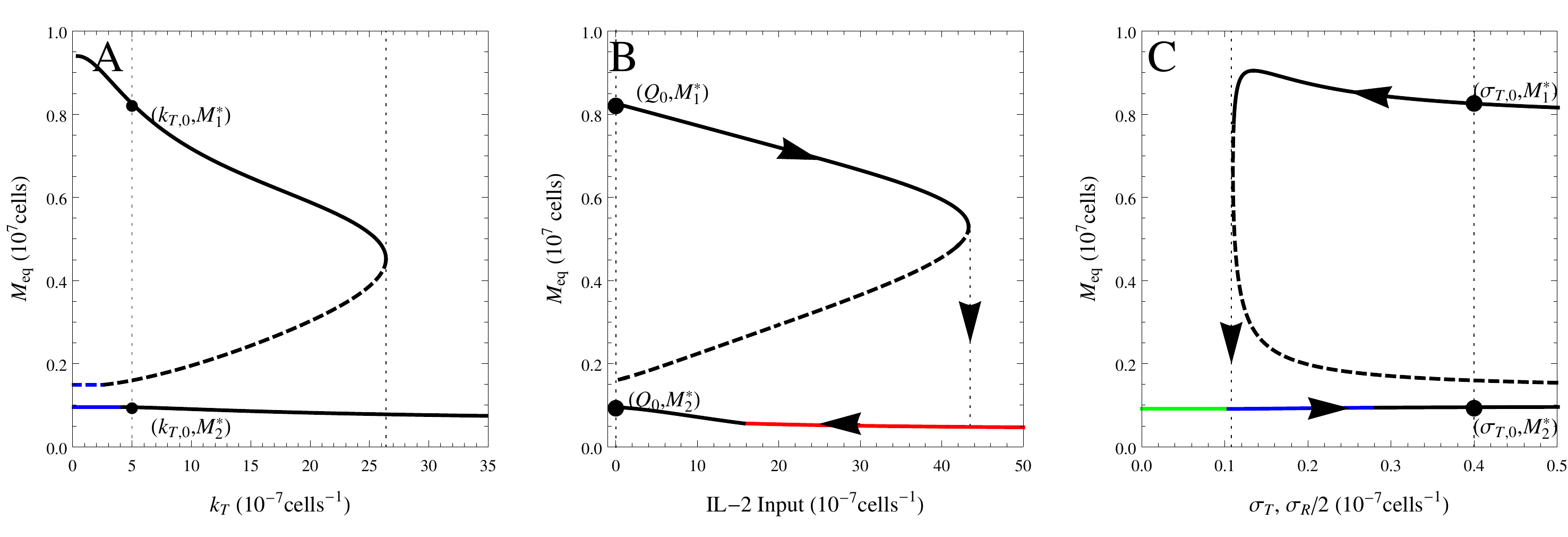}
	\caption{Bifurcation diagrams (A: $k_T$, B: $Q$, and C: $\sigma_T$ and $\sigma_R$ simultaneously) of the initiation model. Solid lines are stable states and dashed
		lines are unstable saddles. Black curves indicate that effectors and Tregs coexist, blue curves have only
effectors, red curves only Tregs, and green curves represent the trivial solution. The arrows indicate the trajectory
during IL-2 treatment (B) and anti-CD3 treatment (C). Dotted lines mark the baseline
values in Table 3.2 and the locations of saddle-node bifurcations.}
	\label{fig:IFNT}
\end{center}
\end{figure}
\end{landscape}

First we vary $k_T$ (Figure \ref{fig:IFNT}A) to simulate different levels of IL-2 production by NOD mice. At the current parameter values (left dotted line),
there are two stable equilibria: severe ($M^*_1$) and mild ($M^*_2$) insulitis. As $k_T$ increases,
the severe equilibrium disappears. This is observed in \cite{sgouroudis2008impact}, when a wild type IL-2 gene bred
into the NOD mice greatly reduces T1D.

Next we increase $Q$ (Figure \ref{fig:IFNT}B) to simulate exogenous IL-2 treatment. The effect is similar to increasing $k_T$, except that the lower
solution branch undergous a transcritical bifurcation (red region) and switches to the $R$-only equilibrium. 
The arrows indicate, conceptually, the trajectory of the system during treatment. The trajectory starts on the upper branch, but is forced
down onto the lower one after the saddle-node bifurcation. After treatment, the trajectory remains on the lower branch.

Finally, we vary $\sigma_R$ and $\sigma_T$ simultaneously while keeping them in the same proportion (Figure \ref{fig:IFNT}C). This represents
treatment with anti-CD3 (which actually decreases these values to zero). The result is almost exactly the same as treatment
with IL-2, with the trajectory dropping down onto the lower branch after the saddle-node bifurcation. Even though this branch
is on the $T$-only equilibrium initially (blue region), it eventually undergoes a transcritical bifurcation to arrive at
the mild insulitis equilibrium.

\subsection{Parameter estimation}
The typical lifespan of a macrophage within the
pancreas is 10 days, so we let $\delta=.1$ \idays \cite{maree2006modelling}. The concentration of these cells in the inflamed pancreas is $1*10^7$cell/ml, 
which we then take to be $J/\delta$, the total number of macrophages in our model at equilibrium \cite{maree2006modelling}.
The macrophage deactivation rate is $b_0=.4$ \cite{maree2006modelling}.
We assume that $\alpha_T+\alpha_R$ has a similiar magnitude to the macrophage influx $J$. From \cite{tang2008central}, we see
that Tregs initially account for 30\% of the CD4 population. Therefore, we let $\alpha_R=.3J$ and $\alpha_T=.7J$. From \cite{ye1995cellular}, we note that IL-12 triggers \IFNg{}
after roughly 8 hours, and so we assume that $a_{\gamma}=c=8$ \idays. Also from \cite{ye1995cellular}, the half life of
the expression of \IFNg{} is about 3 hrs, so $e=5.5$ \idays.

Initially, we let $\nu_{12}=\nu_{10}=\nu_{\gamma}=\delta/J$. This is so that the argument passed to $F$ would be
close to $1$, and so its sigmoidal behavior would be relevant. We find that with these parameters, the basin of attraction
of the severe insulitis state is so small that the equilibrium cannot be reached from reasonable initial conditions. Doubling
the value of $\nu_{\gamma}$ resolves this.

We let $\gamma_T=\gamma_R=2$, which corresponds to a doubling time of 8 hours.
We can directly observe $\sigma_T\approx.4$ and $\sigma_R\approx.8$ from the proliferation data in \cite{tang2008central}.
We also note that, at equilibrium, roughly 10\% of effectors are dividing. Assuming that the proliferation rate matches
the death rate at this point, we find that $\delta_{T1}=.2$. To fit the final parameters, we note that at equilibrium, roughly
20\% of Tregs are dividing and that the ratio of Tregs to effectors is about 9 to 1 \cite{tang2008central}. Taken together,
this allows us to fit the parameters $\delta_{T2}=.8$, $k_T=5\delta/J$ and $k_R=20\delta/J$.

\section{The progression model}
\subsection{CD8 T-cells}
NOD mice have multiple lineages of islet specific T-cells, however, we focus on the high affinity IGRP/NRP-specific
subset. We assume that these cells are generated at a low frequency due to control by central tolerance, but that
they proliferate when the islet antigen IGRP is presented with costimulation in either the Pancreatic Lymph Node or
the islets themselves. Following \cite{mahaffy2007modeling}, we use both a linear death term, representing normal
turnover, and quadratic death term, representing autoregulation by the CD8 population.

\begin{equation}
\frac{dC}{dt}=\alpha_C+\left(\gamma_C \frac{M^*}{M^*+k_C}-\delta_{C1}\right)C-\delta_{C2}C^2
\label{ceqn}
\end{equation}

To parameterize this system we use the data for NRP-A7 CD8 T-cells in \cite{amrani2000progression}. Assuming that
$\alpha_C$ is small and that $M^*$ is constant from week 5 onwards. Then \eqref{ceqn} becomes
\begin{align}
\frac{dC}{dt}&\approx A_i C-\delta_{C2}C^2\\
A_i&=\gamma_C\frac{M^*_i}{M^*_i+k_C}-\delta_{C1}
\end{align}
Where $A_i$ represents the exponential growth rate of CD8 population when $M^*=M^*_i$.
From the time series data, we can estimate that $\delta_{C2}=2.6*10^{-6}$\icells\idays and $A_2=.13$ \idays. 
From \cite{trudeau2003prediction}, we know that there are roughly 8 times as many NRP specific CD8 in mice that get T1D versus those that do not. 
This implies that $A_2\approx8*A_1$. Taken together, these two relations allow us to estimate the values of $\gamma_C=.27$ \idays and $k_C=.75 J$. 
Finally, $\alpha_C$
remains as an important parameter controlling the timing of the expansion of the CD8 T-cells. We adjust
$\alpha_C$, after assigning all other parameters, so that the median onset of T1D occurs at 16 weeks.
This gives $\alpha_C=.03$ cells per day.

\subsection{Metabolic subsystem}
In most experiments, a T1D diagnosis corresponds to a blood glucose level 250mg/dl (normal is 100mg/dl).
Glucose controls both the proliferation rate and insulin production of beta cells. The produced insulin, in turn,
stimulates the uptake of blood glucose. Topp et al.\ \cite{topp2000model} modeled the insulin-glucose system with the following
differential equations.
\begin{align}
\frac{dG}{dt}&=R_0-(E_{G0}-S_I I)G\\
\frac{dI}{dt}&=\sigma_I B \frac{G^2}{G^2+G^2_I}-\delta_I I
\end{align}
As, $\delta_I\gg1$, we assume that the insulin level $I$ is in equilibrium. 
\begin{align}
\frac{dG}{dt}&=R_0-\left(E_{G0}+\frac{S_I \sigma_I}{\delta_I} B \frac{G^2}{G^2+G^2_i}\right)G
\end{align}
To account for the increase in proliferation, we follow the work of \cite{graham2012mathematical} who
created a highly detailed model of beta cell function. We simplify their model as
\begin{align}
\frac{dB}{dt}&=\left(\gamma_B \frac{G^2}{G^2+G^2_{hb}}-\delta_B\right)B.
\end{align}

To parameterize this model we note the following. First, we observe from \cite{sherry2006effects}
that the number of proliferating cells reaches roughly 3\%. 
This corresponds to a max growth rate $\gamma_B=.06$. The lifespan of a typical beta cell is 60 days
\cite{graham2012mathematical} so $\delta_B=1/60$. The resting population of \betacells{} in the absence of an immune response is $B^*=300$,
allowing us to solve for $G_{hb}=161mg/dl$.

In \cite{bleyer2009il2}, untreated mice experience an increase in blood glucose from 300mg/dl to 500mg/dl in the first 2 weeks after onset. 
This corresponds to a decrease in \betacell{} mass from 24 to 4.32 in our model, a loss of roughly 12\% daily. 
To offset compensatory growth, the immune response must remove about 16\% of \betacells{} daily. 
In treated mice, on the other hand, glucose levels drop to about 200mg/dl, which is still twice the baseline level. 
In our model, we can only account for this by a continued immune destruction of about 2\% of immune cells daily. 
This means that the strength of the immune response is roughly 8 times less after treatment than before. 
Interestingly, this prediction corresponds closely to the ratio of NRP-reactive cells found in \cite{trudeau2003prediction} 
between mice that get T1D and those that do not. 
Taken together, we conclude that we can model \betacell{} death due to CD8s as mass action with a constant of $\eta_8=3.2*10^{-6}$ \icells \idays.

CD4s can also kill \betacells{}, although this primarily happens in the absence of Tregs. We include a killing term
due to CD4s, which only becomes relevant to model behavior if Tregs are either absent \cite{chen2005where} or removed
\cite{feuerer2009punctual}. Our final \betacell{} equation is
\begin{align}
\frac{dB}{dt}&=\left(\gamma_B \frac{G^2}{G^2+G^2_{hb}}-\delta_B-\eta_8 C-\eta_4 \frac{(s_{T4}T^*)^2}{1+(s_{T4}T^*)^2+(s_{R4}R)^2}\right)B.
\end{align}

\section{Simulating treatments}
In the next section, we simulate the treatments described in various papers.
Table \ref{dtreats} summarizes the nature and duration of these treatments.
Here we descibe the implementation of IFN-$\alpha$ treatment, anti-CD3 antibodies (aCD3 treatment),
anti-PDL1 antibodies (aPDL1 treatment), Treg treatment, and IL-2 treatment.

\begin{table}[phtb]
\begin{center}
\caption{Summary of Treatments}
\begin{tabular}{lll}
\hline
Treatment & Change & Timeframe\\
\hline
IFN-$\alpha$ Treatment & Day 63-77 & Set $\gamma_C=\nu_{12}=0$\\
Boost of Tregs & Day 77-78 &Increase $\alpha_T$ by $15\delta/J$ \\
\hline
anti-CD3 Treatment & Day 35-49 & Set $\sigma_T=\sigma_R=\delta_{T2}=0$ and $\delta_{T1}=1$\\
anti-PDL1 Treatment & Day 119-133 & Set $\gamma_C=\gamma_T$ and $\nu_{12}=20\delta/J$\\
\hline
IL-2 treatment & Every two days & Pulse $Q$ with $Q_0=500$ and $\delta_Q=1$\idays\\
\hline
\end{tabular}
\label{dtreats}
\end{center}
\end{table}

According to Filippi et al.\ \cite{filippi2009immunoregulatory}, treatment with IFN-$\alpha$ boosts PD-L1 expression. PD-L1 is
a negative costimulatory molecule, so it impairs the ability of APCs
to activate T-cells. We therefore set $\nu_{12}=0$ so CD4s produce no \IFNg{}, and $\gamma_C=0$ so
CD8s do not proliferate \cite{filippi2009immunoregulatory}. Treatment with Tregs is relatively straightforward,
directly increasing the $R$ population, which we implement by changing $\alpha_R$.

Fife et al.\ \cite{fife2006insulin} treat with aPDL1, which we assume should have the opposite effect
to IFN-$\alpha$ treatment. We therefore increase $\nu_{12}$ by a factor of 10 and set $\gamma_C$ equal to
$\gamma_T$, which is the maximum growth rate of T-cells. Anti-CD3 interferes with the interaction between
T-cells and APCs. Thus, we let $\sigma_T=\sigma_R=0$ to indicate that no division takes place. We also
set $\delta_{T2}=0$ and $\delta_{T1}=.1$, which is equivalent to the assumption that both effectors and Tregs die
at the IL-2 deprived death rate of $\delta_{T1}+\delta_{T2}$ as there is no source of IL-2.

Finally, to simulate IL-2 treatment we assume that $Q$ obeys
\begin{align}
	Q&=Q_0\sum_{t_i}e^{(t_i-t)\delta_Q}H(t_i-t)\\
	H(t)&=
\begin{cases}
	t & t>0\\
	0 & \text{else}
\end{cases}
\end{align}
where the $t_i$s are spaced every 2 days for the duration of the treatment. This represents
pulses of $Q_0$ that decay at a rate of $\delta_Q$ per day. This matches the treatment described
in \cite{tang2008central}.

\section{Results}
\subsection{Magnitude of initial inflammation determines\newline T1D prognosis}
\begin{figure}[phtb]
	\includegraphics[width=\textwidth]{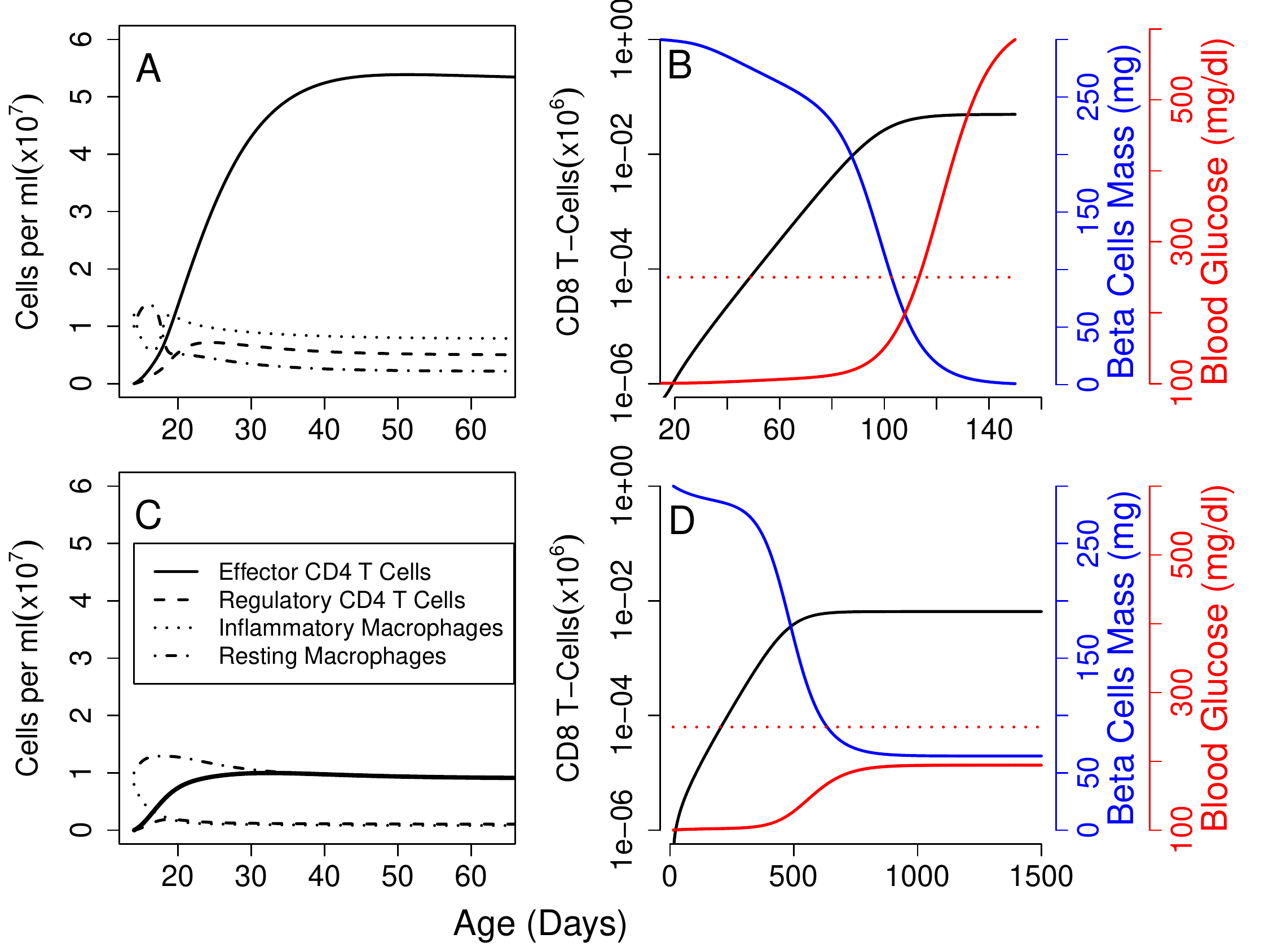}
	\caption{Time series with $M^*(t_0)=1.6*10^7$cells (A,B) or $M^*(t_0)=8*10^6$cells (C,D).
Panels A and C show the CD4 and macrophage populations. Panels B and D show the CD8 population
 (black), \betacells{} (blue), blood glucose (solid red), and the glucose threshold for T1D (250mg/dl, dotted red).}
	\label{fig:tseries}
\end{figure}

As demonstrated in section \ref{analysissection}, the intiation model \eqref{MTeqn} and \eqref{TReqn} has
two stable equilibria, representing insulitis of different severities.
By changing the initial conditions, we can shift the long term behavior of the system from
mild to severe insulitis and the outcome from nondiabetic to diabetic. Specifically, we hold
all initial conditions constant with the exception of activated macrophages. We start each simulation
at $t_0=14$ days, during the apoptotic wave. The initial number of activated macrophages, $M^*(t_0)$, will change depending
on severity of the wave, which could vary between mice. Figure \ref{fig:tseries} shows
time series for $M^*(t_0)=1.6*10^7$cells (Figure \ref{fig:tseries}A,B) and $M^*(t_0)=8*10^6$cells (Figure \ref{fig:tseries}C,D).
When $M^*(t_0)$ is high, the CD4 and macrophage populations equilibriate relatively rapidly
to the severe insulitis state (Figure \ref{fig:tseries}A), eventually leading to T1D onset (Glucose$>$250mg/dl) at 16 weeks
(Figure \ref{fig:tseries}C). When $M^*(t_0)$ is low, the CD4 and macrophage populations equilibriate to the mild insulitis
state (Figure \ref{fig:tseries}C), and T1D does not develop (Figure \ref{fig:tseries}D).

\subsection{Simulation of mouse populations}
One of the main goals of this study is to model the incidence of T1D
in NOD mice under various treatments. Not all NOD mice develop T1D, and the age of onset
can vary among those that do.  As our ODE model is deterministic, we represent the differences between
individuals via parameter values and initial conditions.
To simulate experiments with groups of $N$ mice, we make $N$ parameter sets, sampling each parameter from a
different distribution, described below.
We then replicate those parameter sets and initial conditions for each treatment, so differences between treatments are
never due to stochasticity.
We start each simulation at the time of the initial inflammation ($t_0=14$ days). Therefore, we assume that
the number of activated macrophages will initially be elevated in each mouse, as our model does not include the initiating
event. The initial macrophage population could be different for each mouse due to diet or differences in development. We draw this value from the normal distribution $(1.2+\mathcal{N}(.4,1))*10^7$cells.
Likewise, the value $\eta_8$ is dependent on the affinity of CD8s for \betacells{} and is therefore the outcome of
a complicated process of gene rearrangement, thymic selection, and the population dynamics of competing CD8 clones.
Where specified, we generate this value from the log-normal distribution $3.2*10^{-13}10^{\mathcal{N}(0,1/9)}$\idays.
We generate all other parameters by sampling from a log-normal distribution with the means
given by the base parameters and the standard deviations as 1\% of those means.
When we show the time series from an individual mouse, we use the parameters
in Table \ref{dparameters}.

\subsection{Treatment with IL-2 increases Treg:Teff ratio\newline and prevents T1D}
\begin{figure}[phtb]
	\includegraphics[width=\textwidth]{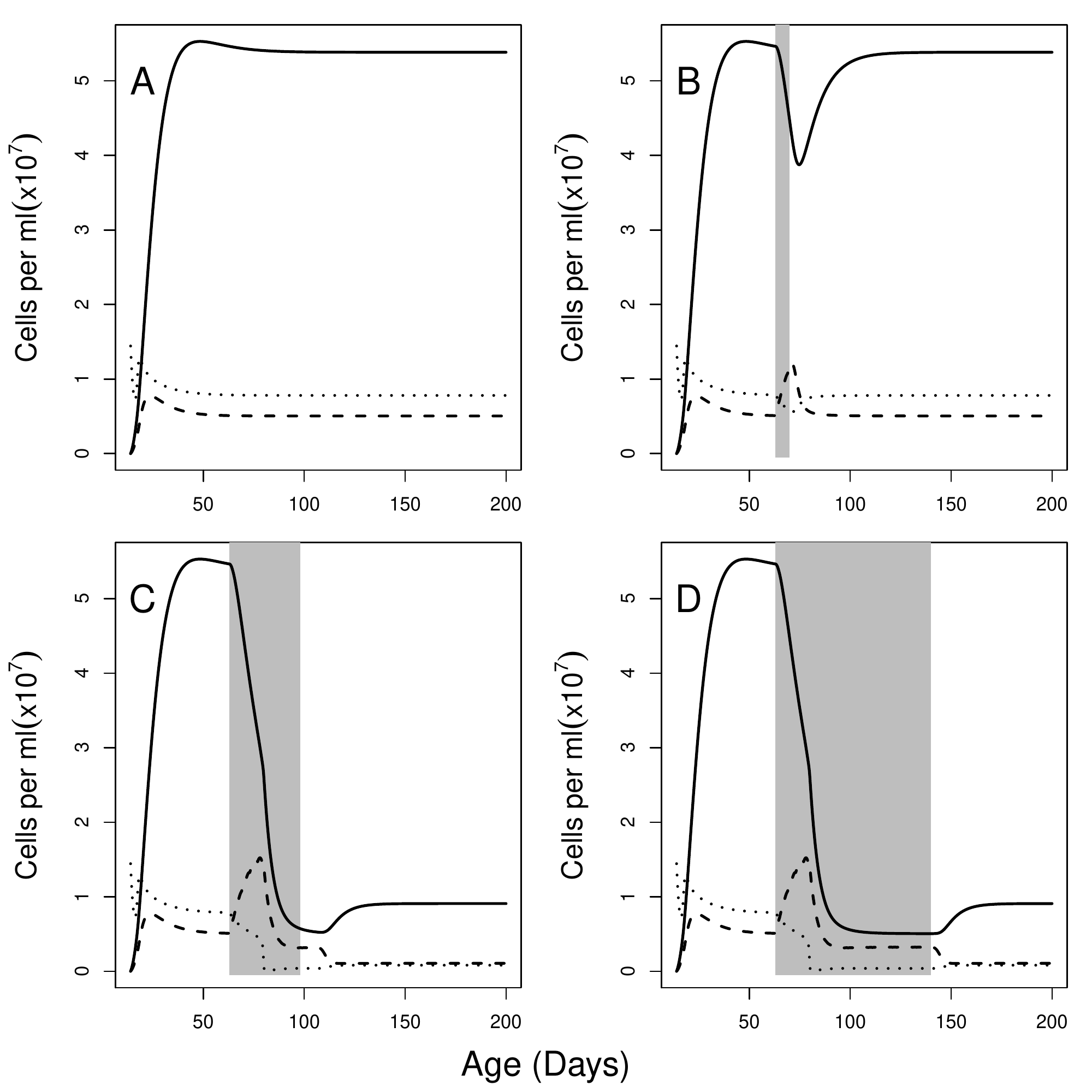}
	\caption{Simulation of IL-2 treatments of different lengths. Using the baseline parameters, we
simulated treatment with IL-2 for 0 (Panel A), 1 (Panel B), 5 (C), or 11(D) weeks. Each panel shows the time course
of the inflammatory macrophages ($M^*$), resting macrophages($M$), effector CD4s ($T$), and Tregs ($R$). The
dosage is $Q=500$.}
	\label{fig:il2}
\end{figure}

\begin{figure}[phtb]
	\includegraphics[width=\textwidth]{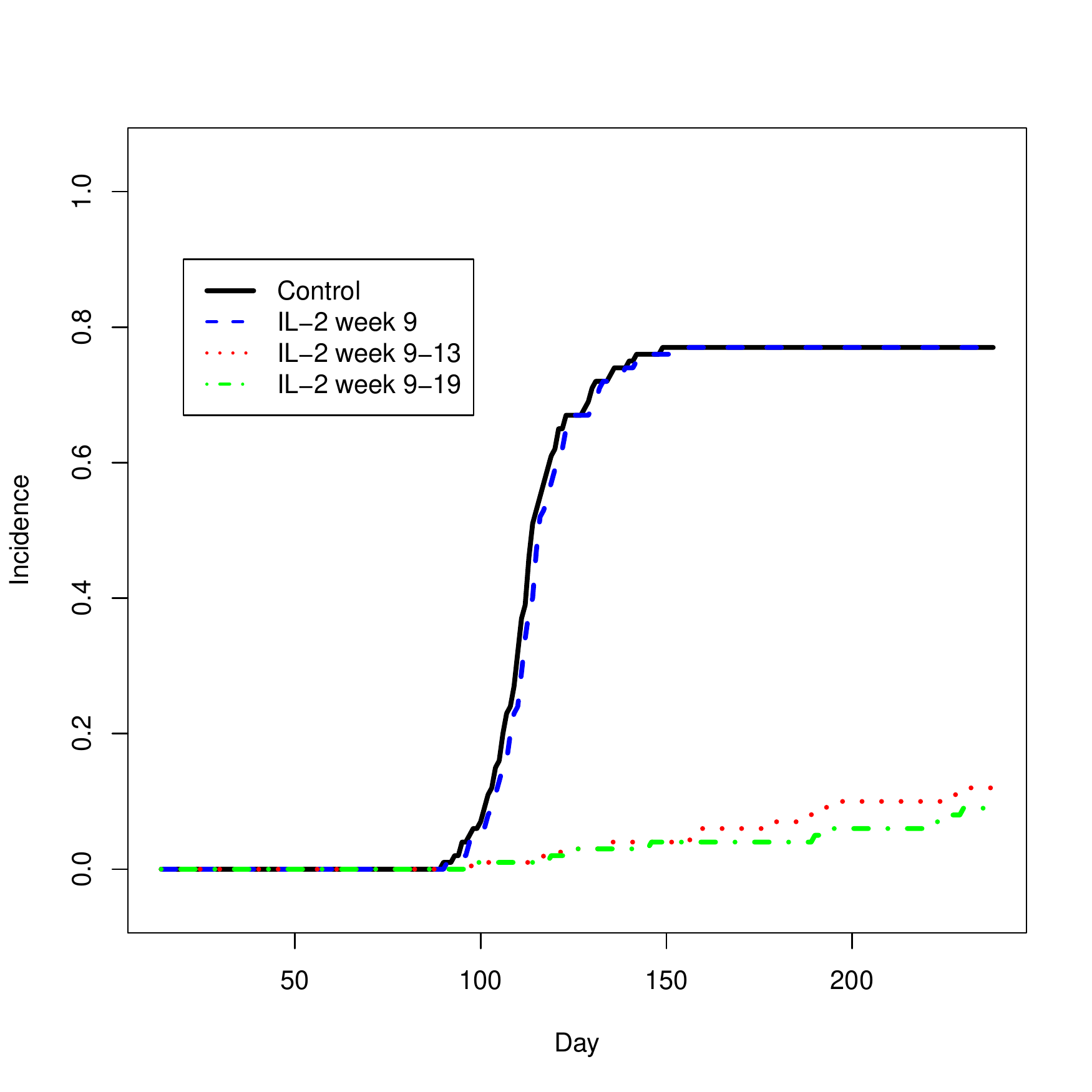}
	\caption{Incidence of T1D with IL-2 treatment for 0, 1, 5, or 11 weeks.}
\label{fig:il2inc}
\end{figure}

Several groups \cite{bleyer2009il2, tang2008central} find that treatment with exogenous IL-2 can restore
the Treg population and prevent the development of T1D. We treat groups of 100 mice with 1, 5, or 11 weeks of IL-2 pulses (of $500$ every 2 days).  Treatment
for 1 week causes a transient but significicant drop in the CD4 population and a corresponding rise in the Treg
population (Figure \ref{fig:il2}A). This results in a very small delay in T1D, but no change in incidence.
Treatment for 5 weeks, on the other hand, leads to a permanent decrease in the CD4 and inflammatory macrophage population
and a permanent increase in the Treg population. This leads to a very large decrease in incidence (Figure \ref{fig:il2inc}). Treatment for 11 weeks has little marginal benefit as compared to 5 weeks. In both of these latter cases, insulitis
is greatly reduced, but remains perpetually. The lower level of insulitis is insufficient, in most cases, to
stimulate the growth of a killer CD8, and so T1D never occurs.

\subsection{CD3 induced tolerance requires continued activity\newline of PD-L1}
Fife et al.\ \cite{fife2006insulin} demonstrate that early treatment with an aCD3 
prevents T1D. CD3 is a surface protein on T-cells that helps them bind to antigen
presenting cells. We can represent CD3 treatment by setting $\sigma_T=\sigma_R=0$ to
indicate that no proliferation takes place. We further modify the death rates of both
effectors and Tregs to be $\delta_{T1}+\delta_{T2}$, indicating that they receive no survival signals
from the IL-2-BCL pathway.
Fife's group also treats some mice with a PD-L1 antibody. PD-L1 acts as a negative regulator of T-cells.
We represent PD-L1 treatment by increasing $\nu_{12}$ 10-fold, corresponding to a decrease
in the activation threshold of CD4 T-cells. This change means that CD4 T-cells require 10 times less
IL-12 to activate and start to produce IFN-$\gamma$. We also assume that more PD-L1 dramatically increases
the division rate of CD8 T-cells to the maximum rate of T-cell division $\gamma_T$.

\begin{figure}[phtb]
	\begin{center}
	\includegraphics[width=\textwidth]{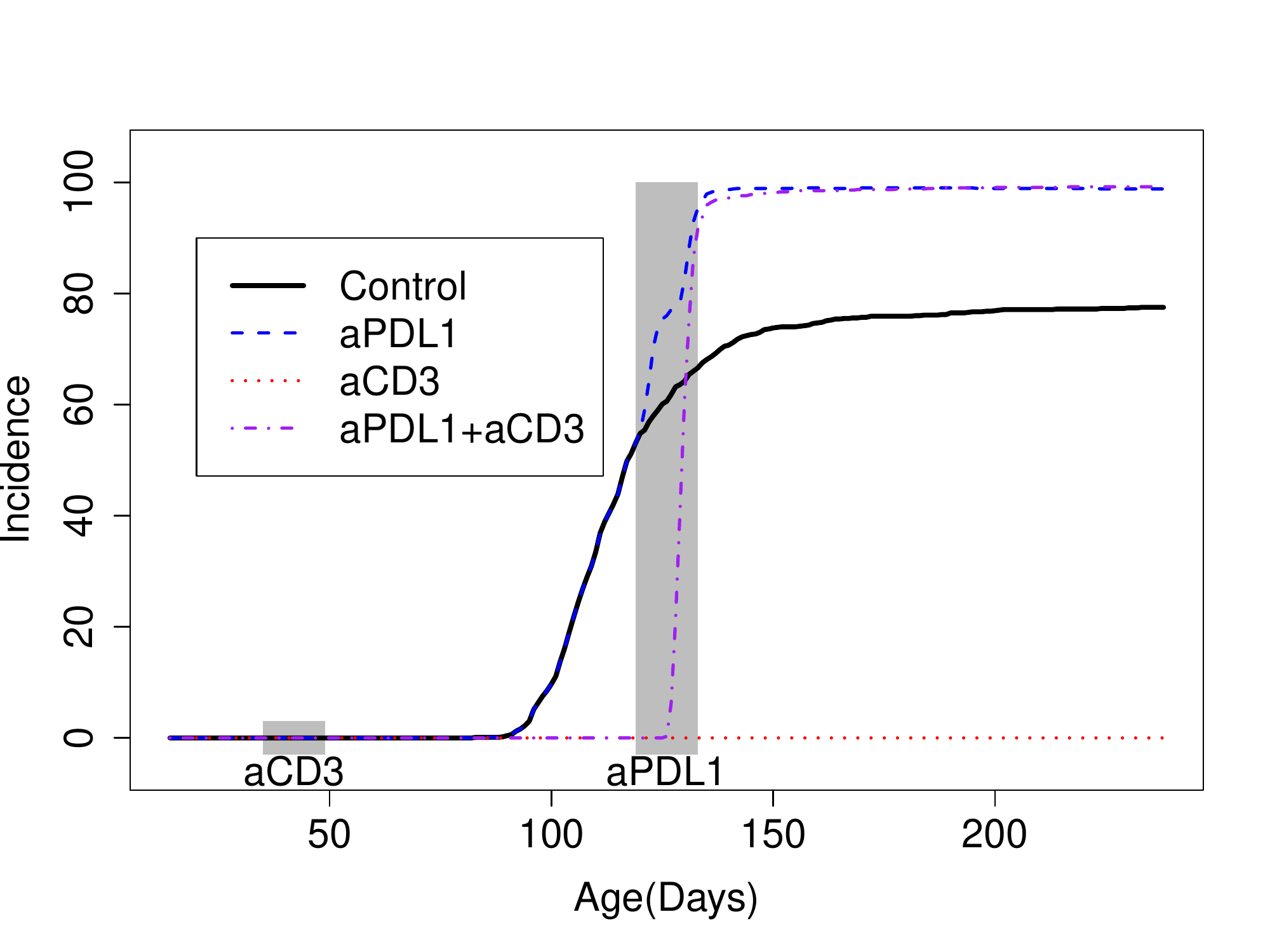}
	\caption{Simulation of treatment of NOD mice with aCD3 and
aPDL1. We generated 100 different parameter sets (`mice') and simulated each T1D
progression under four different simluated treatments. Mice received no treatment, aCD3 at 5 weeks of age,
aPDL1 at 17 weeks, or both treatments. 
This is a replication of experiment from \cite{fife2006insulin}.}
	\label{fig:cd3}
\end{center}
\end{figure}

We follow the same protocol as Fife, treating with aCD3 at 5 weeks and aPD-L1 at 17 weeks, with both treatments
lasting for 2 weeks. Our results (Figure \ref{fig:cd3}) match the key features of the experiment. With aCD3 treatment, aPD-L1 accelerates
T1D among NOD mice and leads to a much higher incidence. Mice that received
aCD3 alone did not develop any T1D, whereas 100\% of those that also received aPD-L1 rapidly developed T1D
shortly after the latter treatment.

\subsection{Synergy between IFN-$\alpha$ and Tregs}
Filippi et al.\ \cite{filippi2009immunoregulatory}
	study the role of viruses in the regulation of  T1D in NOD mice.
They find that the virus LCMV transiently increases PD-L1 expression. In addition,
they hypothesize that the Treg population generated during the immune response
may explain the decreased T1D among LCMV treated mice.
To test this hypothesis, they treat mice with IFN-$\alpha$, which can also increase PD-L1
expression and transfer Tregs from mice previously exposed to LCMV.

We represent the boost in PD-L1 by setting $\nu_{12}=0$, meaning that APCs cannot activate
CD4 T-cells to produce IFN-$\gamma$, and $\gamma_C=0$, meaning that CD8 T-cells cannot divide.
This is esentially the inverse of how we modeled the aPD-L1 treatment. We represent the
injection of T-cells by increasing $\alpha_R$ for the duration of the treatment.

\begin{figure}[phtb]
	\begin{center}
	\includegraphics[width=\textwidth]{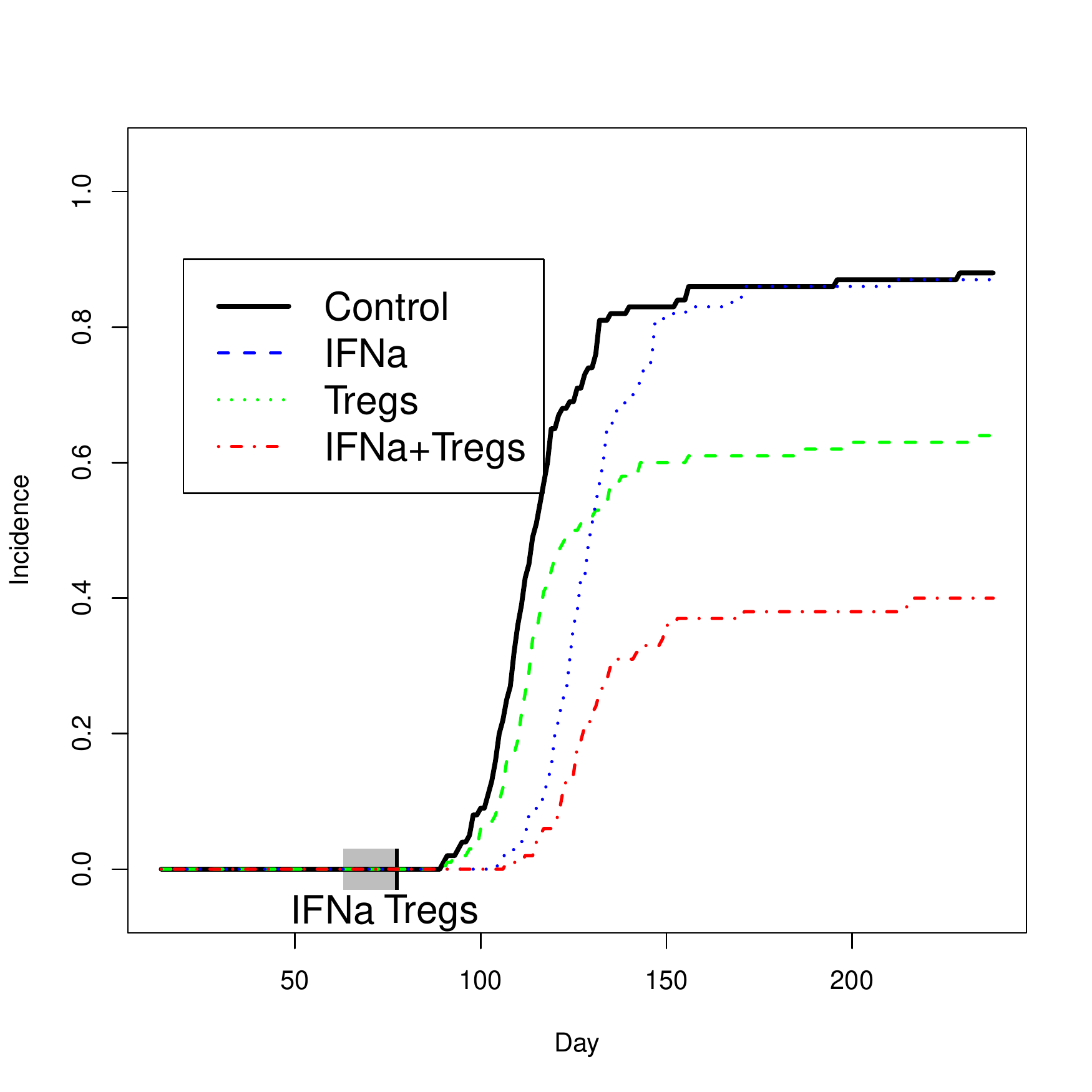}
	\caption{Simulation of treatment of NOD mice with IFN-$\alpha$
		and Tregs. We generated 100 different parameter sets (`mice') and simulated each T1D
progression under four different simluated treatments. 
Mice received no treatment, IFN-$\alpha$ at 9 weeks of age,
Tregs at 11 weeks or both treatments. 
	This is a replication of experiment from \cite{filippi2009immunoregulatory}.}
	\label{fig:double}
\end{center}
\end{figure}
Like Fillipi, we find synergy between the two treatments (Figure \ref{fig:double}). The administration of IFN-$\alpha$ by itself does
not change the incidence of T1D, but it does delay the age of onset. The administration of Tregs by themselves
does not change the age of onset, but does decrease incidence. The combination of both treatments decreases
incidence by more than the sum of the individual treatments. All of these observations replicate Fillipi's findings.
In this experiment, we used a smaller range of $\eta_8$ to bring it in line with the other parameters. Without this change,
the variance is so large that it obscures the synergy.

\subsection{CD4 and macrophage parameters affect incidence while\newline CD8 parameters affect age of onset}
To determine which parameters contribute to incidence and which to age of onset, we ran a sensitivity analysis for each parameter. 
First we generate 100 pairs of values of $\eta_8$
and $M^*(t_0)$, our two key parameters.
Then we vary each parameter, one at a time, from 50\% to 200\% of its baseline value, equally spaced on
a log scale, with every other parameter fixed at its baseline. For each value of the current parameter,
we simulate the system using each of the 100 parameters pairs for $\eta_8$ and $M^*(t_0)$. We record if and when
T1D develops in each simulation. 

\begin{figure}[phtb]
	\includegraphics[width=\textwidth]{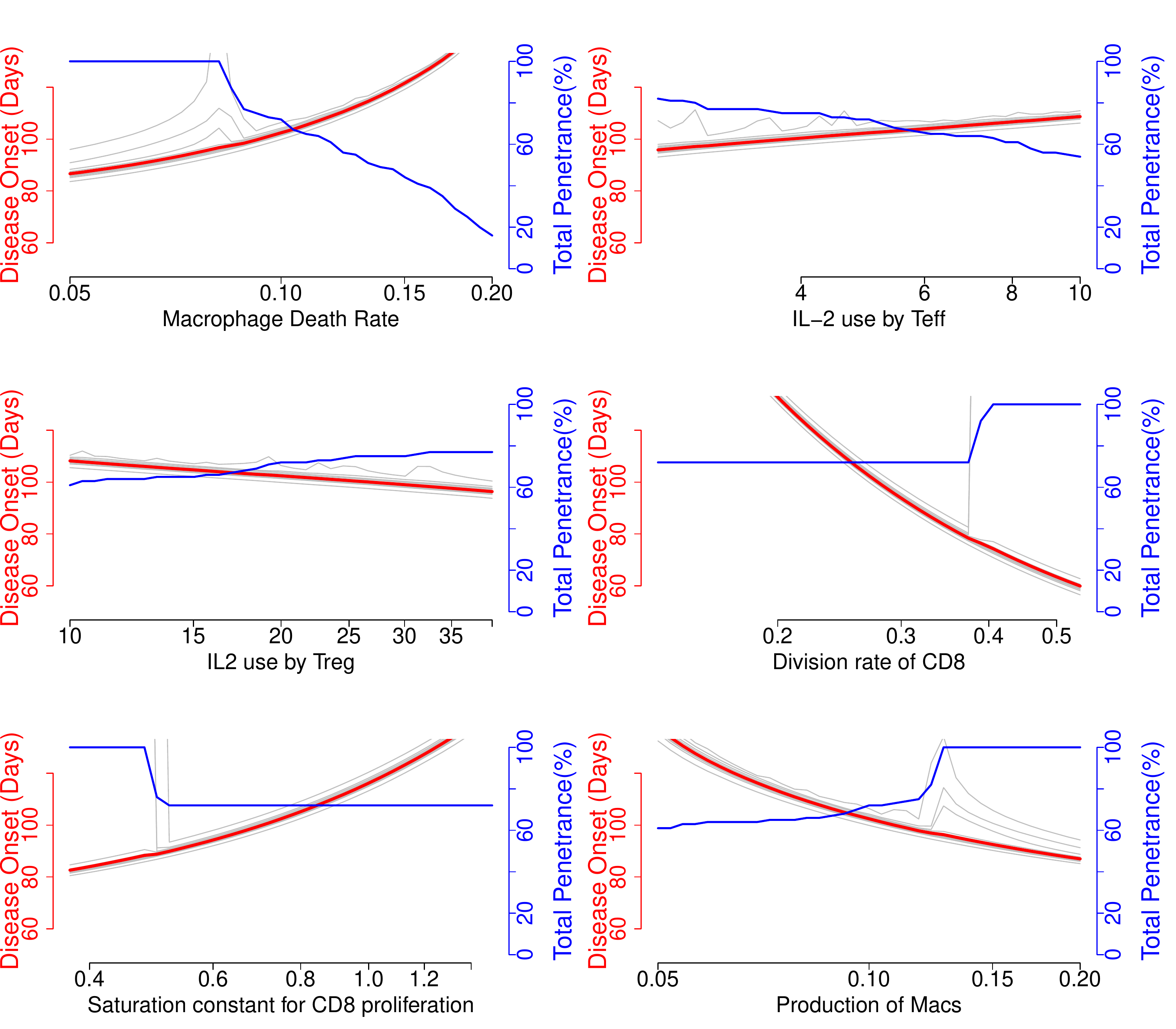}
	\caption{Sensitivity analysis of full model. Parameters that primarily affect age of onset. We vary an individual parameter
	from 50\% to 200\%
        of its baseline value. We hold each other parameter constant except for the killing
rate of CD8 $\eta_8$ and the initial number of activated macrophages $M^*(t_0)$.
 	The blue line shows the percentage of diabetic mice, by the end of 50 weeks, in the sample.
 	The red and grey lines show the median and deciles, respectively, of the age of onset.}
	\label{fig:sa1}
\end{figure}

\begin{figure}[phtb]
	\includegraphics[width=\textwidth]{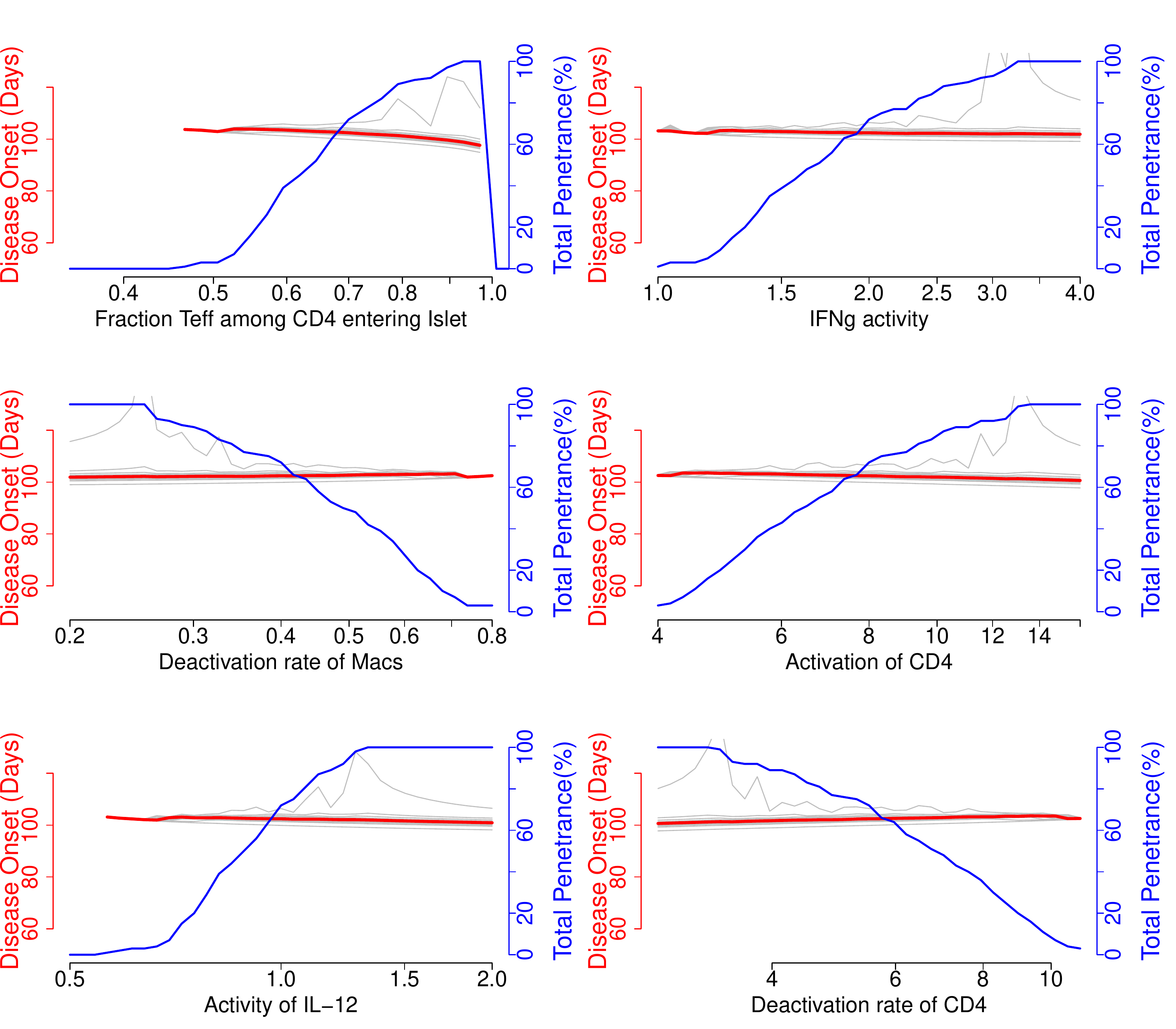}
	\caption{Sensitivity analysis of full model. Parameters that primarily affect incidnece. We vary an individual parameter
	from 50\% to 200\%
        of its baseline value. For each other parameter, we randomly sample 100 times from
	a normal distribution centered on the baseline value and run a seperate simulation for
each parameter set.
 	The blue line shows the percentage of diabetic mice, by the end of 50 weeks, in the sample. 
 	The red and grey lines show the median and deciles, respectively, of the age of onset.}
	\label{fig:sa2}
\end{figure}

\begin{figure}[phtb]
	\includegraphics[width=\textwidth]{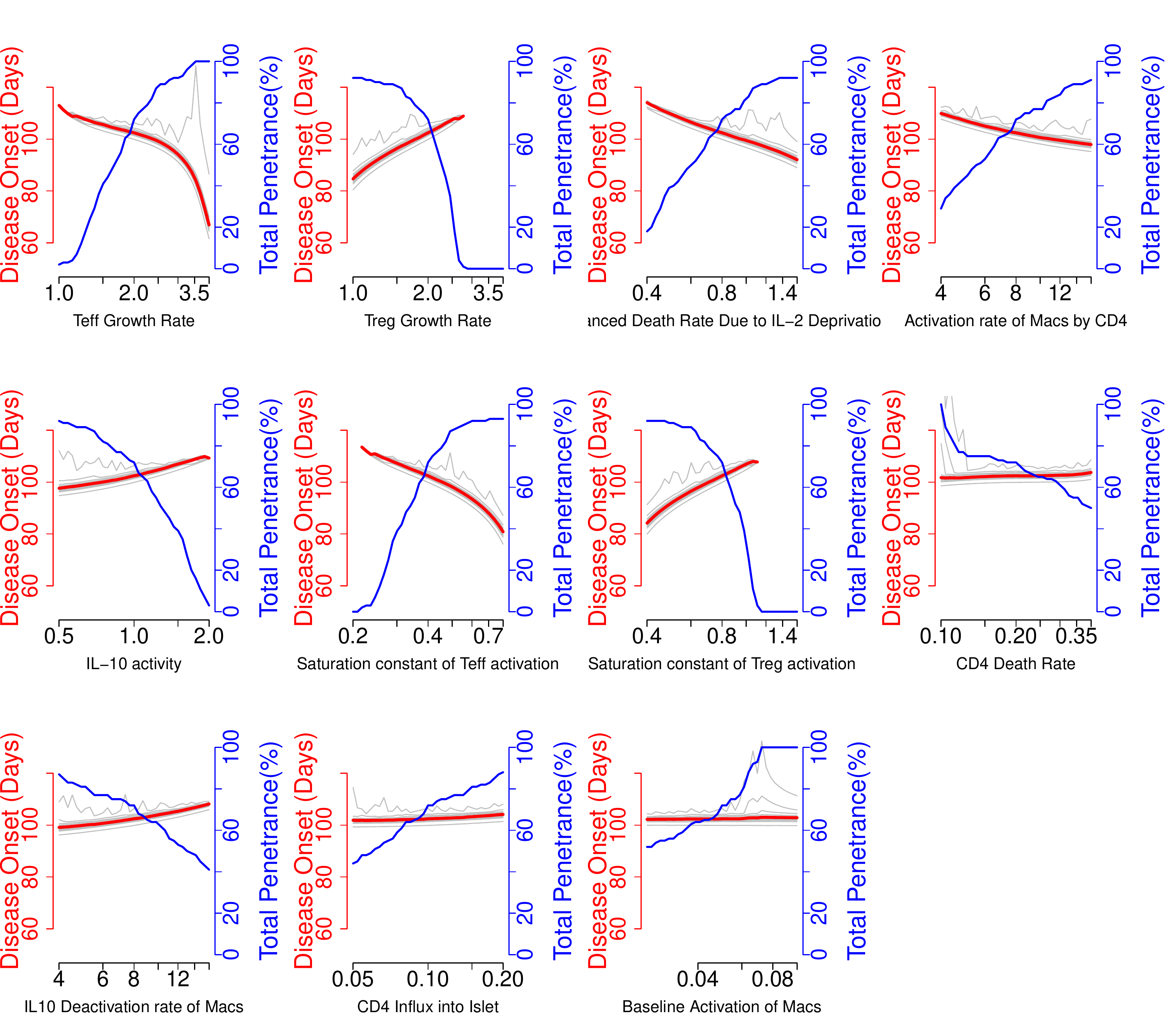}
	\caption{Sensitivity analysis of full model. Parameters that affect incidence and age of onset. We vary an individual parameter
	from 50\% to 200\%
        of its baseline value. For each other parameter, we randomly sample 100 times from
	a normal distribution centered on the baseline value and run a seperate simulation for
each parameter set.
 	The blue line shows the percentage of diabetic mice, by the end of 50 weeks, in the sample.
 	The red and grey lines show the median and deciles, respectively, of the age of onset.}
	\label{fig:sa3}
\end{figure}

\begin{figure}[phtb]
	\includegraphics[width=\textwidth]{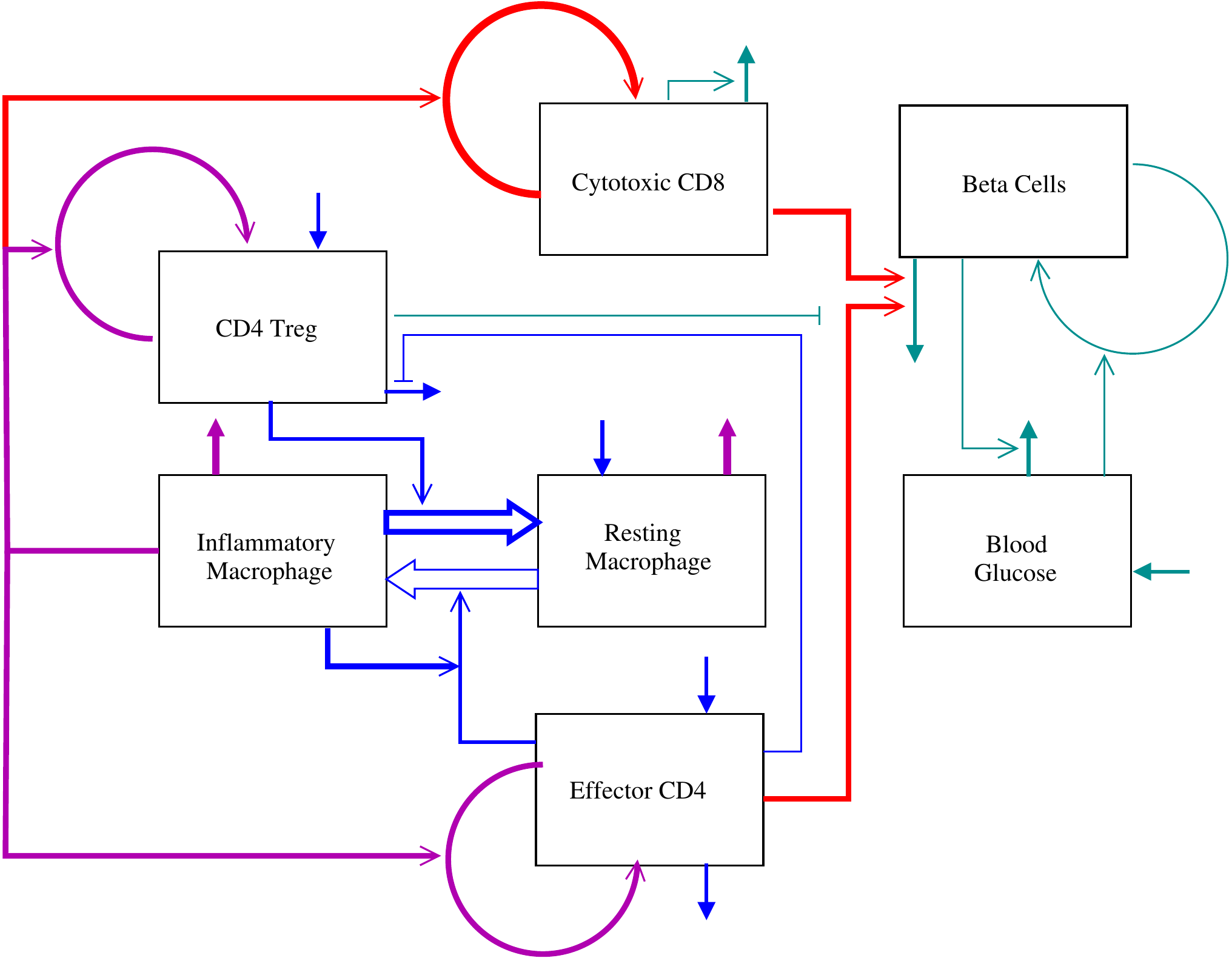}
	\caption{Schematic diagram summarizing the model in this paper. Arrows are
	colored according to whether they affect incidence/penetrance (blue), age of onset (red),
	or both (purple). The strength of the effect is denoted by the arrow thickness. Arrows in dark cyan
	had negligible effect.}
	\label{fig:diabmodelfig}
\end{figure}

Figures \ref{fig:sa1}, \ref{fig:sa2}, and \ref{fig:sa3} summarize the results
of the sensitivity analysis. In each panel, we vary a different parameter. The blue line shows the incidence, while
the red and grey lines show the median and deciles, respectively, of the age of onset. Figure \ref{fig:sa1} shows
parameters that primarily affect age of onset, Figure \ref{fig:sa2} shows those that primarily affect incidence, and
Figure \ref{fig:sa3} shows those with strongs effects on both. In general, the parameters that describe CD4 and macrophage
behavior affect only incidence and those that describe CD8 behavior affect only the age of onset. We summarize this pattern
in Figure \ref{fig:diabmodelfig} that shows the entire model, with each arrow color-coded and scaled according to its significance.
Bolder arrows have greater significance, red arrows affect primarily age of onset, blue affect incidence, and purple
arrows affect both. Greyed-out arrows have little affect on the model and are confined to the metabolic processes in the model.

\section{Discussion}
In this paper, we present a mathematical model of Type 1 diabetes (T1D) in the NOD mouse. We propose
that the intensity of the apoptotic wave controls the eventual development of T1D. We further propose
that the long delay between this initial inflammation and the destruction of the islets is due to the growth and maturation of the CD8 population. Therefore, the model
has two components: an `initiation' component that consists of equations governing the interaction of CD4s and macrophage
populations, and a `progression' component that describes the growth of CD8s, their killing of \betacells{}, and the eventual
rise in blood glucose. 
The initiation component has two possible outcomes, defined by its stable equilibria, only one of which leads to T1D. These
equilibria are distinguished by the relative activity of Tregs, effectors, and activated macrophages and their associated cytokines: IL-10, IFN-$\gamma$ and IL-12.

Our model reproduces the results of several experiments on NOD mice: aCD3 treatment,
aPDL1 treatment, IL-2 treatment, IFN-$\alpha$ treatment,
and Treg treatment. All of these treatments amount to shifting the trajectory from the basin of attraction
of the severe-insulitis state to that of the mild insulitis state. A treatment that does not cause a transition can
still cause a delay in insulitis. For example, IFN-$\alpha$ treatment alone does not decrease incidence but does
delay T1D due to downstream effects that interfere with the CD8 population. When combined with adoptive transfer of Tregs, 
it does significantly decrease incidence. According to our model, each treatment by itself is insufficient to make
the transition, and so the system returns to the severe-insulitis state once treatment is over. Conversely, treatment with
aPDL1 can shift the system from mild to severe arthritis. We show that aCD3 treatment followed by aPDL1 shifts
the system from severe to mild and back to severe again. The equivalence between all of these treatments suggests that the
system is `memoryless'. A mouse that has been cured with aCD3 should be similar to one cured via the transfer of Tregs
or IFN-$\alpha$. Fife et al.\ \cite{fife2006insulin} found similarities between aCD3 treated mice and those that received
insulin coupled splenocytes that tolerized them to the peptide. Table 3.4 summarizes the experimental results that
our model can reproduce.

In our model, Tregs play two separate roles. First, Tregs prevent the killing of \betacells{} by CD4 T-cells. This is an
assumption of the model given that Tregs have been observed to control the extent of infiltration in the islet 
\cite{chen2005where, feuerer2009punctual}. Second, Tregs can control the inflammatory state of macrophages and other APCs which in turn prevents the development of a CD8 response.
 Due to IL-2 deficiency in NOD mice, Tregs are at a competitive disadvantage in the islets, favoring a proinflammatory environment.
In our model, this proinflammatory envinronment is manifested in the severe insulitis equilibrium. We find that a moderate increase in IL-2 production can eliminate this equilibrium.
Treatment with exogenous IL-2 does not eliminate
this equilibrium in the long term, but can shift the system to the mild equilibrium as the other treatments do.

\begin{table}[phtb]
\begin{center}
\caption{Key behaviors this model can reproduce}
\begin{tabular}{lll}
	\hline
Phenomenom & Source & Outcome\\
\hline
T1D at 16 weeks & \cite{alanentalo2010quantification} & Fitted\\
T1D in 80\% of females & \cite{anderson2005nod}& Fitted\\
T1D rapid without Tregs & \cite{chen2005where} and \cite{feuerer2009punctual}&\\
Protection from T1D with enhanced IL-2 production&\cite{sgouroudis2008impact}&\\
Protection from T1D with exogenous IL-2&\cite{sgouroudis2008impact}&\\
Protection from T1D with anti-CD3 treatment & \cite{fife2006insulin}&\\
Protection from T1D with Treg treatment&\cite{filippi2009immunoregulatory}&\\
Delay of T1D with IFN-$\alpha$ treatment&\cite{filippi2009immunoregulatory}&\\
Synergy between Treg and IFN-$\alpha$ treatments&\cite{filippi2009immunoregulatory}&\\
NOD8.3 have faster onset, same incidence &\cite{verdaguer1997spontaneous}&\\
IGRP-specific CD8 predict outcome&\cite{trudeau2003prediction}&\\
Inflammation predicts T1D months in advance&\cite{fu2012early}&\\
\hline
\end{tabular}
\label{behav}
\end{center}
\end{table}
It has been suggested that the Treg population loses either effectiveness \cite{tritt2008functional} or population size \cite{tang2008central} over time. Lack
of IL-2 uptake by Tregs decreases BCL-2 expression and thus survival. Therefore, the IL-2 deficiency lowers the Treg:effector ratio in 
NOD islets relative to the spleen and lymph nodes and relative to wild type islets. 
However, this ratio does not decline as the mouse ages. Bleyer et al.\ find that it stays constant until disease onset, 
when it \textit{rises} slightly \cite{bleyer2009il2}. In our model, the Treg:effector ratio stays constant over the progression of T1D and is not
significantly different between mice that develop T1D and those that do not. This is not to say that the Treg population
does not decline in NOD mice, but we need not assume it to reproduce the known phenomena. The presence of T1D-resistant
mice, either naturally or following treatment by IL-2, IFN-$\alpha$, or aCD3, suggests ongoing regulation of the insulitic legion
by Tregs. If the Tregs were to intrinsically decline, we would expect all of these mice to also develop T1D. Although this
process may happen on a timescale longer than current experiments, it cannot be the driving force behind onset of T1D.

CD8 killer T-cells drive the eventual decline of the \betacell{} population. Although, in this model, CD4 T-cells also possess 
the capacity to destroy \betacells, they do not do so in the presence of Tregs (as in \cite{feuerer2009punctual}).
The CD8 population in our model represents the population of high affinity CD8 specific to IGRP. Trudeau et al.\ \cite{trudeau2003prediction} use
the size of this population to predict T1D outcome, and Amrani et al.\ \cite{amrani2000progression} show that an increase in 
the average affinity of the CD8 population for IGRP correspond with disease onset. 
To eliminate the \betacell{} population, the CD8 population must kill them faster than the \betacells{} can divide. We estimate that at the time of disease onset, CD8 T-cells
kill \betacells{} at a rate of roughly 16\% a day, which is almost 3 times as fast as \betacells{} have been observed
to divide. Therefore, a slight decrease in CD8 number or effectiveness is unlikely to prevent T1D development, although it
can slow it. We find that CD8 related parameters primarily contribute to the age of onset instead of T1D incidence. In
agreement with this, TCR8.3 NOD transgenic mice that produce only high-affinity IGRP-specific CD8 T-cells have a more rapid onset, but similiar
incidence to NOD mice \cite{verdaguer1997spontaneous}.

In this model, the treatment of the CD8 population is extremely simple. We only track a single population that grows from a very small number of precursors.
In NOD mice, the CD8 population transitions from insulin specific
to IGRP specific \cite{amrani2000progression}. An initial response to insulin is required for T1D progression
\cite{krishnamurthy2006responses}, but it is unclear whether this is due specifically to the activity of CD8s.
It is plausible that initial low levels of \betacell{} death due to insulin-specific CD8s releases IGRP, which is then
presented on APC triggering the switch in autoimmunity. It is also plausible that the CD8-mediated death is
necessary to maintain the `severe-insulitis' state that leads to T1D in this model. One possible experiment to elucidate
the role of CD8s would be to create a transgenic line of mice on the NOD background whose CD8s lack the ability to kill \betacells. According to
our simple model of CD8s, they should still develop an IGRP-specific population in the same time frame. If these NOD
mice continue to produce primarily insulin-specific CD8s, that will imply that CD8-related pathology drives the progression
of their own \betacell{} affinity.

Another simplifying assumption of our model is that the \betacell{} population responds only to glucose levels
and only homeostatically.
In fact, \betacells{} divide in response to islet inflammation \cite{sreenan1999increased}. We have ignored this phenomenon, reasoning that
the immune response results in a net decrease in \betacells{}. In addition, \betacells{} can decompensate \cite{sreenan1999increased}
when under high demand, producing less insulin per cell. This is due in part to the degranulation of \betacells{} during
T1D progression \cite{akirav2008beta}. Including any of these phenomena will change the parameter estimates for the immune-mediated
killing of \betacells{}. More importantly, it could alter some of the conclusions of how we expect the \betacell{} population
to respond to treatment. For example, the inclusion of degranulated \betacells{} could allow for a rapid rebound after
IL-2 treatment in new onset mice, as in \cite{bleyer2009il2}, which cannot be reproduced by the one-compartment \betacell{}
model we use.

In treating T1D in both NOD mice and humans, there are two main strategies. Administration of anti-inflammatories, such
as anti-CD3, Vitamin D or omega-3s \cite{jayasimhan2014advances} can reduce insulitis and hopefully prevent the development
of an autoimmune destruction. Antigen specific tolerance, such as with oral administration of insulin,
aims to delete the autoreactive T-cell clones directly \cite{jayasimhan2014advances, zhang2008insulin}.
Only the former have had successful trials among humans, but the latter represent a
more specific treatment.

\bibliographystyle{plain}
\bibliography{autoimmune}
\end{document}